\documentclass[twocolumn,showpacs,preprintnumbers,amsmath,amssymb]{revtex4}
\usepackage{epsfig}
\usepackage{amsmath}

\begin{document}

\title{Transient and Equilibrium Synchronization in Complex Neuronal Networks}

\author{Luciano da Fontoura Costa}
\affiliation{Institute of Physics at S\~ao Carlos, University of
S\~ao Paulo, PO Box 369, S\~ao Carlos, S\~ao Paulo, 13560-970 Brazil}

\date{16th Jan 2008}

\begin{abstract}
Transient and equilibrium synchronizations in complex neuronal networks
as a consequence of dynamics induced by having sources placed at
specific neurons are investigated.  The basic integrate-and-fire neuron
is adopted, and the dynamics is estimated computationally so as to
obtain the activation at each node along each instant of time.  In the
transient case, the dynamics is implemented so as to conserve the
total activation entering the system.  In our equilibrium
investigations, the internally stored activation is limited to the
value of the respective threshold. The synchronization of the
activation of the network is then quantified in terms of its
normalized entropy.  The equilibrium investigations involve the
application of a number of complementary characterization methods,
including spectra and Principal Component Analysis, as well as of an
equivalent model capable of reproducing both the transient and
equilibrium dynamics.  The potential of such concepts and measurements
is explored with respect to several theoretical models, as well as for
the neuronal network of \emph{C. elegans}.  A series of interesting
results are obtained and discussed, including the fact that all models
led to a transient period of synchronization, whose specific features
depend on the topological structures of the networks.  The
investigations of the equilibrium dynamics revealed a series of
remarkable insights, including the relationship between spiking
oscillations and the hierarchical structure of the networks and the
identification of twin correlation patterns between node degree and
total activation, implying that hubs of connectivity are also hubs of
integrate-and-fire activation.
\end{abstract}

\maketitle

\vspace{0.5cm}
\emph{`Measure what is measurable, and make measurable what is not so.' 
(G. Galilei)}

\section{Introduction} 

Few research areas have received so much attention as neuronal
networks and complex networks.  Investigations at the intersection of
these two areas are particularly promising because they naturally
integrate the emphasis on structure typical of complex networks and
the dynamics systems features of neuronal networks.  Such a
multidisciplinary area is henceforth called \emph{complex neuronal
networks}.  Despite the many possibilities for cross-fertilization
between these two areas, relatively few related works have been
reported.  Probably the first investigations in neuronal complex
networks were reported in~\cite{Stauffer_Hopfield, Stauffer_Costa}.
Other related works include~\cite{Costa_revneur:2005, Kim:2004,
Timme:2006}.  Another issue of special interest regarding the dynamics
of systems underlain by complex connectivity regards their
synchronization.  Synchronization plays an important role in several
natural situations, including heartbeats and overall brain activity.
Several works addressing synchronization in complex networks have been
reported in the literature (e.g.~\cite{Watts_Strogatz:1998,
Watts:2003, Watts:2004, Boccaletti:2006,
Hong_etal:2004,Lee:2005,Boccaletti:2005, Zhou:2006, Arenas:2008,
Boccaletti:2007, Lodato:2007, Takashi:2006, Sorrentino:2007, Ott:2007,
Almendral:2007, Costa_sync:2008}). The investigation of
synchronization in neuroscience and neuronal networks has a long
history (e.g.~\cite{Borisyuk:1998, Aoki:2004,Percha:2005,
Pereira:2007}).  More recently, growing attention has been placed on
synchronization of complex neuronal networks (e.g.~\cite{Osipov:2007,
Hasegawa:2004,Hasegawa:2005,Park:2006}).  However, a great deal of
such investigations has had their attention concentrated on models of
individual neuronal activation (e.g. Hodgkin-Huxley or Kuramoto) which
are more sophisticated that the integrate-and-fire individual neuron
model.  Though many interesting results have been produced by such
works, it is felt that a more systematic exploration of the
relationship between synchronization and the different types of
topologies in complex networks can be achieved by using simpler models
for the neuronal dynamics, such as the integrate-and-fire approach,
which allows faster computational simulations

The current article addresses the study of synchronization during the
transient activation of complex networks with different topological
features involving neurons of the integrate-and-fire type under
conservation of the incoming activity.  Although intrinsically related
to neuronal networks, such results extend immediately to other types
of systems involving integration-and-fire dynamics, especially
production systems in which the output of a node requires the
integration of specific inputs in order to yield a product
(e.g. production of molecules, industrial production, or computational
calculations in parallel systems).  In this respect, the delay
implemented by the combination of the integrator and non-linear
element would act as a mechanism for favoring the chances of receiving
all required inputs before the node can act on them.  

Specific nodes are chosen as source of activation, and their effect in
activating other neurons is sound and objectively quantified in terms
of the instantaneous synchronization, a measurement defined in terms
of the normalized entropy of the node activations at a given instant
of time.  The normalized instantaneous synchronization (NIS) is
related to the entropy of the system activation
(e.g.~\cite{Latora_entropy, Gardenes:2007, Costa_sync:2008}), but also
takes into account the number of nodes already activated.  Such an
analysis of the overall synchronization of the network is performed
with respect to 6 theoretical models of networks (Erd\H{o}s-R\'enyi,
Barab\'asi-Albert, Watts-Strogatz, a geographical network, as well as
two knitted networks).  The neuronal network of
\emph{C. elegans}~\cite{Watts_Strogatz:1998} is also investigated.  The
obtained results indicate that the overall synchronization tends to
present a peak along the medium term, collapsing subsequently.  In
addition, each type of network implied specific features regarding the
network activation and instantaneous synchronization.  

The patterns of regular spiking arising at the equilibrium as a
consequence of synchronization of the individual firings are also
investigated by using a series of characterization methodologies as
well as an equivalent model.

This article starts by summarizing the main concepts in complex
networks and neuronal networks and proceeds by describing the
normalized instantaneous synchronization concept, which is
subsequently applied for the characterization of the synchronization
in 6 networks of distinct kinds.  The equilibrium investigations are
reported subsequently with respect to Erd\H{o}-R\'enyi and
Watts-Strogatz Networks.

\section{Basic Concepts}

This section summarizes the main concepts in network representation,
measurement, random walks, as well as the six network models assumed
in the present article.

\subsection{Complex Networks Basics and Models}

A weighted, directed network $\Gamma$ can be fully represented in
terms of its \emph{weight matrix} $W$.  Each edge extending from node
$i$ to node $j$, with associated weight $v$, implies $W(j,i) = v$.
The absence of connection between nodes $i$ and $j$ implies
$W(j,i)=0$.  The \emph{out-degree} of a node $i$, henceforth expressed
as $k_{out}$, corresponds to the number of outgoing edges of that
node.  The \emph{out-strength} of a node $i$, $s_{out}$, is given by
the sum of the respective weights of all outgoing edges.  Similar
definitions hold for the \emph{in-degree} and \emph{in-strength}.

Six models of complex networks are considered in the present article:
Erd\H{o}s-R\'enyi (ER), Barab\'asi-Albert (BA), Watts-Strogatz (WS), a
geographical model (GG)~\cite{Albert_Barab:2002, Newman:2003,
Dorogov_Mendes:2002}, as well as the path-regular network (PN) and
path-transformed model (PA) ~\cite{Costa_surv:2007, Costa_path:2007,
Costa_comp:2007, Costa_longest:2007}.  The ER network (see also
related works by~\cite{Flory}) was grown by taking each possible edge
with constant probability, the BA structure was grown by using the
traditional preferential attachment scheme~\cite{Albert_Barab:2002},
and the WS was derived from a linear regular network with rewiring
rate 0.1~\cite{Watts:2003,Watts:2004}.  The geographical network (GG)
is obtained by distributing the nodes through a two-dimensional space
and connecting all pairs of nodes which have distance smaller than a
fixed threshold.  The two knitted networks are the path-regular (PN)
and path-transformed BA (PA)
networks~\cite{Costa_path:2007,Costa_comp:2007}.  Both these networks
are formed by paths.  The PN network is grown by incorporating paths
involving all network nodes, being intensely regular regarding several
of its topological features~\cite{Costa_comp:2007,
Costa_longest:2007}.  The PA network can be obtained by transforming
(from stars to paths) a BA network with the same number of nodes.  All
networks in this work have similar number of nodes and average degree.
Only the largest connected component has been taken into account for
each network.  However, because of the relatively high average degree
adotped in this work ($\left< k \right> = 6$), most of the nodes end
up belonging to the largest component.

\subsection{Hierarchical Organization of Complex Networks}

Given a complex network and a ~\emph{reference node} $i$, its
~\emph{hierarchical organization} can be obtained by flooding the
network from the reference node~\cite{Costa:2004}.  The hierarchical
organization includes the ~\emph{concentric levels} (or hierarchical
levels) of the network, namely the levels containing the nodes which
are at successive shortest path distances from the reference node $i$.
So, the first concentric level incorporates the original nodes which
are at shortest path distance 1 from $i$ (i.e. they are the immediate
neighbors of $i$).  The second concentric level includes the nodes
which are at shortest path distance 2 from $i$, and so on.  The
connections between the reference node and the nodes at the $h-$level
can be understood as implementing ~\emph{virtual
links}~\cite{Costa_perc:2004}.  Once the hierarchical organization of
a network has been obtained with respect to a specific node chosen as
the reference, a series of measurements can be
calculated~\cite{Costa:2004, Costa_NJP:2007, Costa_JSP:2006,
Costa_EPJB:2005}, including the ~\emph{hierarchical number of nodes}
$n_h(i)$, the ~\emph{hierarchical degree} $k_h(i)$ and the
~\emph{intra-ring degree} $a_h(i)$.  The hierarchical number of nodes
corresponds to the number of nodes within each respective concentric
level.  The hierarchical degree is equal to the number of edges from
level $h$ to level $h+1$.  The intra-ring (or intra-level) degree
corresponds to the number of edges established within level $h$.

With the extension of the hierarchical organization to complex
networks with asymmetric connections~\cite{Costa_eqcomm:2008}, the
hierarchical degree needs to be split into ~\emph{hierarchical
indegree} and ~\emph{hierarchical outdegree}.  Therefore, the
hierarchical indegree of level $h$ with respect to node $i$,
henceforth abbreviated as $ki_h(i)$, is equal to the number of edges
received by level $h$ from level $h-1$ .  The hierarchical outdegree
of level $h$ is the number of edges sent from that level to level
$h+1$.

\subsection{Instantaneous synchronization}

Let $\Sigma$ be a dynamic system implemented over a complex network
involving $N$ nodes.  The activation of each node $i$ at each time $t$
is henceforth represented as $A(i,t)$.  For simplicity, and without
loss of generality, such activations can be normalized so that they
become a statistical distribution.  This can be done by defining the
\emph{probability of activation} of node $i$ at time instant $t$ to be
$a(i,t) = A(i,t) / \sum_{i=1}^{N}A(i,t)$.  The \emph{entropy} of all
such activations at $t$ (e.g.~\cite{Latora_entropy,Gardenes:2007}) can
not be immediately given as

\begin{equation}
  \epsilon(t) = - \sum_{i=1}^{N} A(i,t) log(A(i,t))
\end{equation}

Note that the maximum value of the entropy, corresponding to $log(N)$,
is achieved when all nodes have the same probability of activation
$1/N$, i.e. the total activation is the most uniformly distributed
amongst all the nodes.  

Because the activation probability can be understood as a \emph{mean
frequency of activation} (e.g.~\cite{Costa_sync:2008}), it is
interesting to consider the \emph{instantaneous synchronization} of
the system.  A possibility is to use the following
expression~\cite{Costa_sync:2008}

\begin{equation}
  \sigma(t) = \frac{log(N) - \epsilon(t)}{log(N)}
\end{equation}

Note that $0 \leq \sigma(t) \leq 1$, with the maximum entropy leading
to null synchronization and minimal entropy leading to maximum
synchronization.  The nodes with null activation are not considered in
the calculation of the entropy used for the synchronization because
they are not really participating to the overall activation dynamics. 

Though such a definition properly reflects the relationship between
activation entropy and instantaneous synchronization, it does not take
into account the fact that, especially during the transient period of
time (but sometimes also at steady state), some nodes will not be
active.  Let us illustrate this problem through the following example.
Let a dynamical system with $N=100$ nodes have only 2 nodes activated
at time $t$, e.g. because we are in the transient period and
activation has not yet reached the other nodes.  The probability
activation therefore will be $a(t)=0.5$, yielding $\epsilon(t) \approx
0.69$ and $\sigma = (log(100) - \epsilon(t)) / log(100) \approx 0.85$,
indicating a high level of synchronization between the two nodes.
Although this is really the case when only the two nodes are
considered, the system actually involves other 98 nodes which are at
zero activation.  In order to better express the overall
synchronization considering all the $N$ nodes, we adopt henceforth the
following alternative definition of the instantaneous synchronization
of the dynamics:

\begin{equation}
  \xi(t) = \frac{N_a}{N} \sigma(t) = \frac{N_a(t)}{N} \frac{log(N) -
  \epsilon(t)}{log(N)} \label{eq:xi}
\end{equation}

where $N_a(t)$ is the number of nodes with non-zero activity at time
$t$.  Now, we will only have maximum synchronization $\xi(t)=1$ when
$N_a(t)=N$ and all nodes have the same activation probability.  This
measurement, henceforth called \emph{normalized instantaneous
synchronization} (NIS), is adopted throughout this work.  Going back
to the previous example, we now have $\xi(t) \approx (0.85)(0.2)
\approx 0.017$, which provides a more reasonable quantification of the
instantaneous synchronization considering the whole network.

\subsection{Spectral Characterization}

Given a time series or signal $s(t)$, it is often quite difficult to
identify its periodical components.  By reinforcing the intrinsic
periodicities along the signal, its \emph{autocorrelation} allows a
more effective means for inferring its constituent oscillations.  In
this work we resource to the \emph{power spectrum} $P(f)$ of the
signal $s(t)$, which corresponds to the squared magnitude of the
Fourier transform of the autocorrelation of
$s(t)$~\cite{Costa_book:2001}.  More specifically, given the signal
$s(t)$, its Fourier transform is defined as

\begin{eqnarray}
  S(f) = \int_{t=-\infty}^{\infty} s(t) exp(-i2\pi ft)  dt  \nonumber \\ \nonumber
\end{eqnarray}

The autocorrelation of $s(t)$ is

\begin{eqnarray}
  a(\tau) = \int_{t=-\infty}^{\infty} s(t) s(\tau-t)  dt  \nonumber \\ \nonumber
\end{eqnarray}

Therefore, the Fourier transform of the autocorrelation function is
immediately given by the correlation theorem as

\begin{eqnarray}
  A(f) = S(f) S(f)^{*}   \nonumber \\ \nonumber
\end{eqnarray}

where `$S(f)^*$' is the conjugate of S.  The \emph{magnitude of the
spectrum} of $s(t)$ is 

\begin{eqnarray}
  |A(f)| = \sqrt{S(f) S(f)^{*}}  dt  \nonumber \\ \nonumber
\end{eqnarray}

The \emph{power spectrum} of $s(t)$ is defined as being equal to the
squared magnitude of the spectrum of $s(t)$, i.e.

\begin{eqnarray}
  P(f) = S(f) S(f)^{*}   \nonumber \\ \nonumber
\end{eqnarray}

The power spectrum allows the identification of autocorrelations
(especially oscillations) of the original signal.  More specifically,
presence of peaks at specific frequencies $f$ in the power spectrum
indicate the presence of respective oscillations with frequency $f$ in
the original signal $s(t)$.  All power spectra in this work are
estimated by using the Fast Fourier Transform (FFT).

\subsection{Multivariate Statistical Methods}

The spikes produced along time for each neuron in the complex neuronal
networks can be understood as \emph{patterns}, which can be compared
and classified by using multivariate
statistics~\cite{McLachlan:1998,Duda_Hart:2000, Costa_book:2001}
and/or pattern recognition methods~\cite{Duda_Hart:2000,
Costa_book:2001}.  In the present work, we apply the Principal
Component Analysis (PCA) methodology~\cite{Costa_surv:2007,
Costa_book:2001} in order to decorrelate the spike patterns
~\cite{Nicolelis:1998} and to obtain more significant clusters in
respective projections of the original patterns.

Given a set of $M$ spike patterns along $H$ time steps, a total of $H$
measurements corresponding to the presence of a spike at each time can
be obtained.  Let us define the matrix $U$ so that each of its rows
corresponds to a train of spikes, so that $U$ has dimension $M \times
H$.  Let $\vec{\mu}$ be the $1 \times H$ vector containing the average
number of spikes at each time step $h = 1, 2, \ldots, H$.  Now, define
the new matrix $F$ as

\begin{eqnarray}
  F = U - ones(M,1) \vec{\mu} \nonumber \\ \nonumber
\end{eqnarray}

where $ones(M,1)$ is a $M \times 1$ vector of ones.  The
\emph{covariance matrix}, taking into account all pairwise covariances,
is immediately given as

\begin{eqnarray}
  C = \frac{1}{M-1} F F^T \nonumber \\ \nonumber
\end{eqnarray}

Observe that $C$ is symmetric.  Let $\lambda_i$ be the eigenvalues of
$C$ sorted in descending order, with respective eigenvectors
$\vec{v_i}$.  The Karhunen-Lo\`eve transform of the original
measurements is the stochastic linear transformation implemented by
the matrix G given as follows

\begin{equation} \label{eq:PCA}
  G  =  \left [ \begin{array}{ccc}
              \longleftarrow  &  \vec{v_1}  &  \longrightarrow  \\
              \longleftarrow  &  \vec{v_2}  &  \longrightarrow  \\
              \ldots          &  \ldots     &  \ldots  \\
              \longleftarrow  &  \vec{v_m}  &  \longrightarrow  \\
             \end{array}   \right]  
\end{equation}

The PCA method involves transforming the original measurements as $V =
G U^T$ with $m << H$, as allowed by the high redundancy normally found
along and between signals.  It can be shown that the resulting new
measurements $V$, which correspond to linear combinations of the
original measurements, are completely decorrelated by the PCA
methodology, therefore maximizing the variation of the data long the
first new variables.

\section{Modeling and Simulation}

\begin{figure}[htb]
  \vspace{0.3cm} \begin{center}
  \includegraphics[width=1\linewidth]{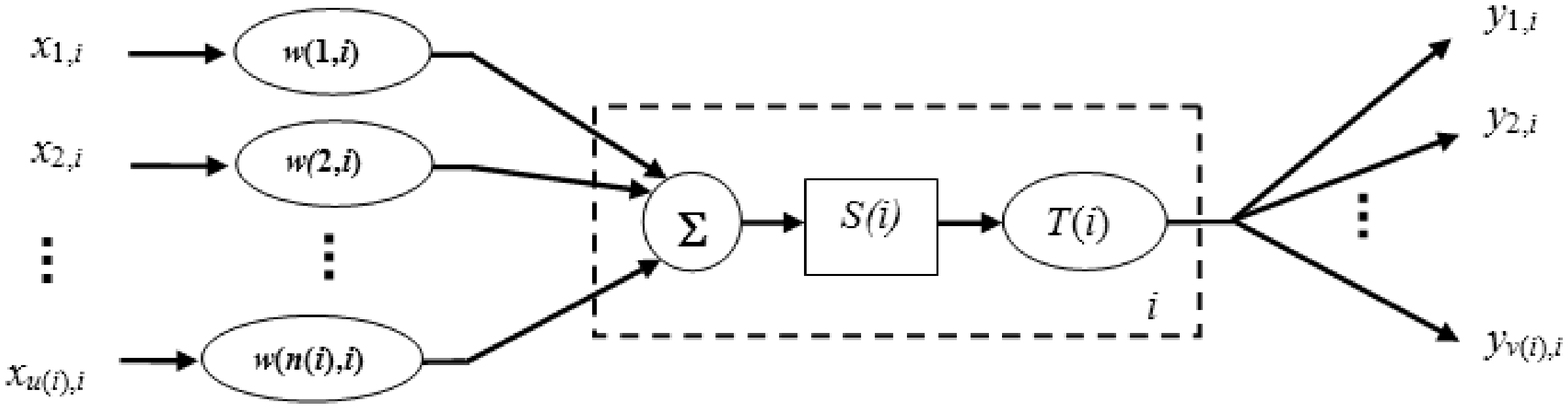} 
  \caption{The integrate-and-fire type of neuron adopted in this work.
  }~\label{fig:neuron} \end{center}
\end{figure}

In this work, we focus attention on the two following specific
features: (i) integrate-and-fire dynamics at each node; and (ii)
conservation of incoming activation.  Figure~\ref{fig:neuron} shows
the basic node adopted henceforth, which corresponds to a simple
integrate-and-fire neuron.  Each such node $i$ includes $n(i)$ inputs
and $m(i)$ outputs.  The input activity is integrated until its value
reaches the threshold $T(i)$ (hard limit non-linearity is adopted in
this work), in which case the neuron fires.

The integrator and non-linear stages are henceforth understood to
constitute the \emph{soma} (or body) of the neuron.  The above type of
neuron can be immediately modified in order to allow the conservation
of the total activation which has already entered the system (the
system may involve sources of activation).  First, once the neuron
fires, all its store activation is liberated through the outgoing edge
(i.e. axons).  Observe that the total liberated activation will be
necessarily equal to the threshold $T(i)$, and that the activation
will be divided amongst the outgoing edges, so that each receive a
fraction of activation equal to $y_{i,j} = 1/k_{out}(i)$ (so,
$k_out(i) = m(i)$).  Second, the total activation received by the
neuron is stored internally until the neuron fires. After each spike
and transfer of the currently stored activation, the state of the
neuron is cleared. These two simple modifications allow the activation
which has already entered the system to be completely conserved along
time.  It should be observed that the division of activation amongst
the outgoing connections is not biologically reasonable because the
action potential in neurons involves signal reinforcement (the axon is
active) and is known to produce spikes with similar intensities in
most synapses.  However, a biologically realistic network can be
easily set up in order to reproduce the conservative dynamics by
making the synapses at which the axons $y_{i,j}$ terminate to have
weights $1/k_{out}(i)$.  

In the case of the equilibrium analyses, the activation inside each
memory $S(i)$ is limited to $L(i)$ at all times.  This implies that
the portion of the activation received by a node which exceeds $L(i)$
is discarded, therefore undermining the conservation of the activation
received from the source node after the neurons start firing.

The complex neuronal networks considered in this article consist of a
representative sample of each of the ER, BA, WS, GG, PN and PA models.
The dynamics is implemented by considering each of the nodes in these
networks to be a conservative integrate-and-fire neuron as discussed
above.  All neurons are assumed to have the same threshold $T =
1$. The activation of the network is implemented by assigning a source
to a specified node, which therefore acts as a source of constant
activation with intensity $1$.  As time passes, such an activation is
distributed to the other nodes in the networks.  The total of
activated neurons at any time $t$ is $N_a(t)$.  In order to ensure
activation conservation, the weight of each connection from node $i$ to
node $j$ is defined as $w(j,i) = 1/k_{out}(i)$.  For the sake of
simplicity, each of the undirected edges yielded by the 6 considered
network models are dissociated into one dendrite and one axon, so that
the out-degree becomes identical to the in-degree.  Less-symmetric
configurations can be considered futurely.  The \emph{C. elegans}
network is kept directed in our simulations.

The dynamics of such complex neuronal networks has been investigated
with respect to: (a) the evolution of the activity of all nodes along
time, represented in diagrams which are henceforth referred to as
\emph{activograms}; (b) the distribution of activated and non-activated 
nodes along time; (c) the distribution of the spikes produced by all
neurons along time (\emph{spikegram}); and (d) the evolution of the
normalized instantaneous synchronization along time.  The maximum NIS
values obtained while considering the source at every node, as well as
the time at which such values occur, are also considered in this work.

\section{Transient Synchronization: Results and Discussion}

Simulations on the theoretical models were performed considering
$n=100$ and $\left< k \right> = 6$.  The largest connected component
in the neuronal network of \emph{C. elegans} contained 239 neurons.
For both the theoretical and real-world networks, each of the nodes
was considered as a source of activation with intensity 1.

\begin{figure}[htb]
  \vspace{0.3cm} \begin{center}
  \includegraphics[width=1\linewidth]{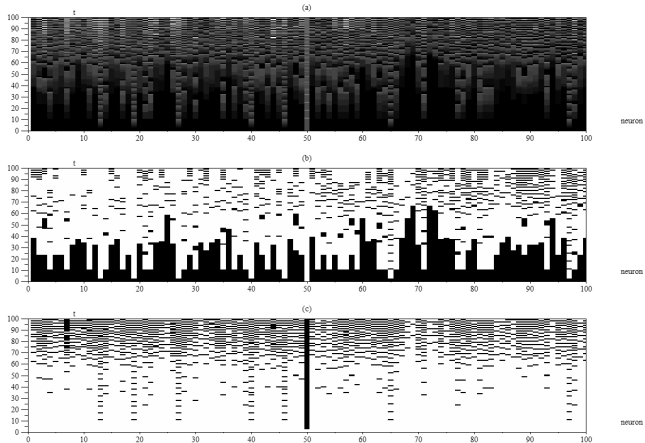}
   \caption{Activation diagrams for ER network: (a) probability
  activation of each node along time (a); the active and non-activate neurons
  along time (b); and the spikes produced by each neuron
  along time (c).  The source was placed at node 50.
  }~\label{fig:grams_ER} \end{center}
\end{figure}

\begin{figure}[htb]
  \vspace{0.3cm} \begin{center}
  \includegraphics[width=1\linewidth]{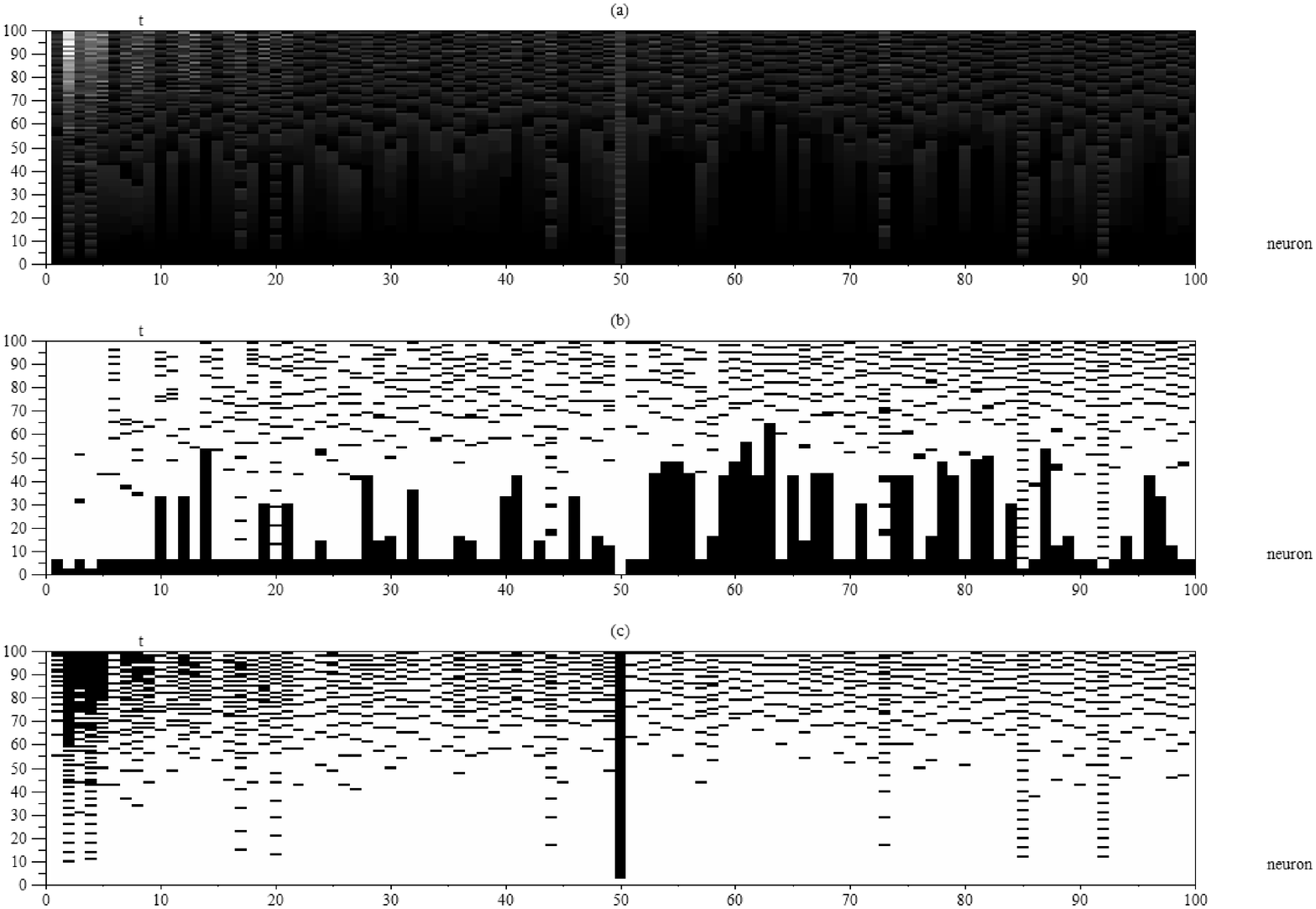}
   \caption{Activation diagrams for BA network: (a) probability
  activation of each node along time (a); the active and non-activate neurons
  along time, shown in black (b); and the spikes produced by each neuron
  along time (c).  The source was placed at node 50.
  }~\label{fig:grams_BA} \end{center}
\end{figure}

\begin{figure}[htb]
  \vspace{0.3cm} \begin{center}
  \includegraphics[width=1\linewidth]{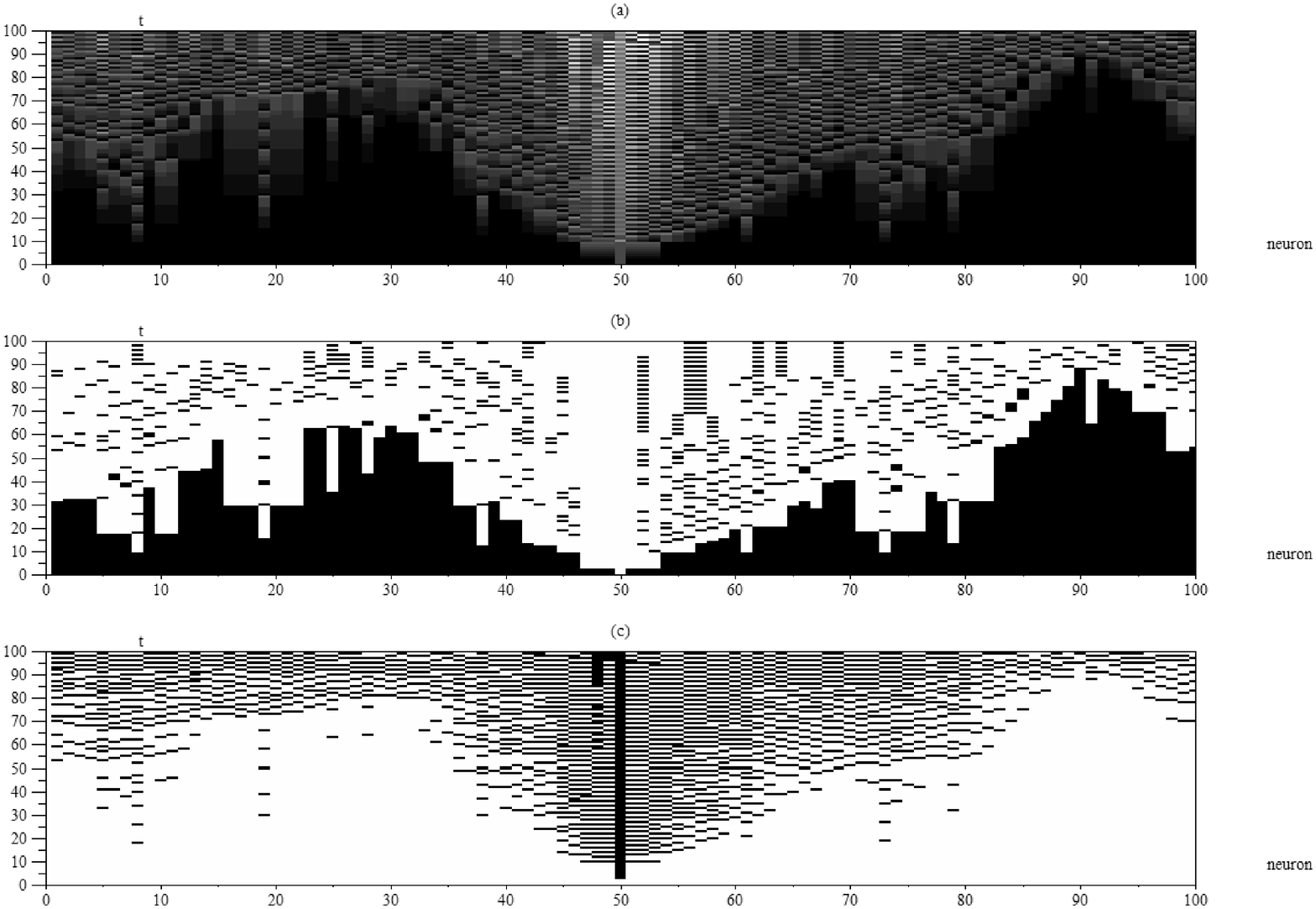}
   \caption{Activation diagrams for WS network: (a) probability
  activation of each node along time (a); the active and non-activate neurons
  along time (b); and the spikes produced by each neuron
  along time (c).  The source was placed at node 50.
  }~\label{fig:grams_WS} \end{center}
\end{figure}

\begin{figure}[htb]
  \vspace{0.3cm} \begin{center}
  \includegraphics[width=1\linewidth]{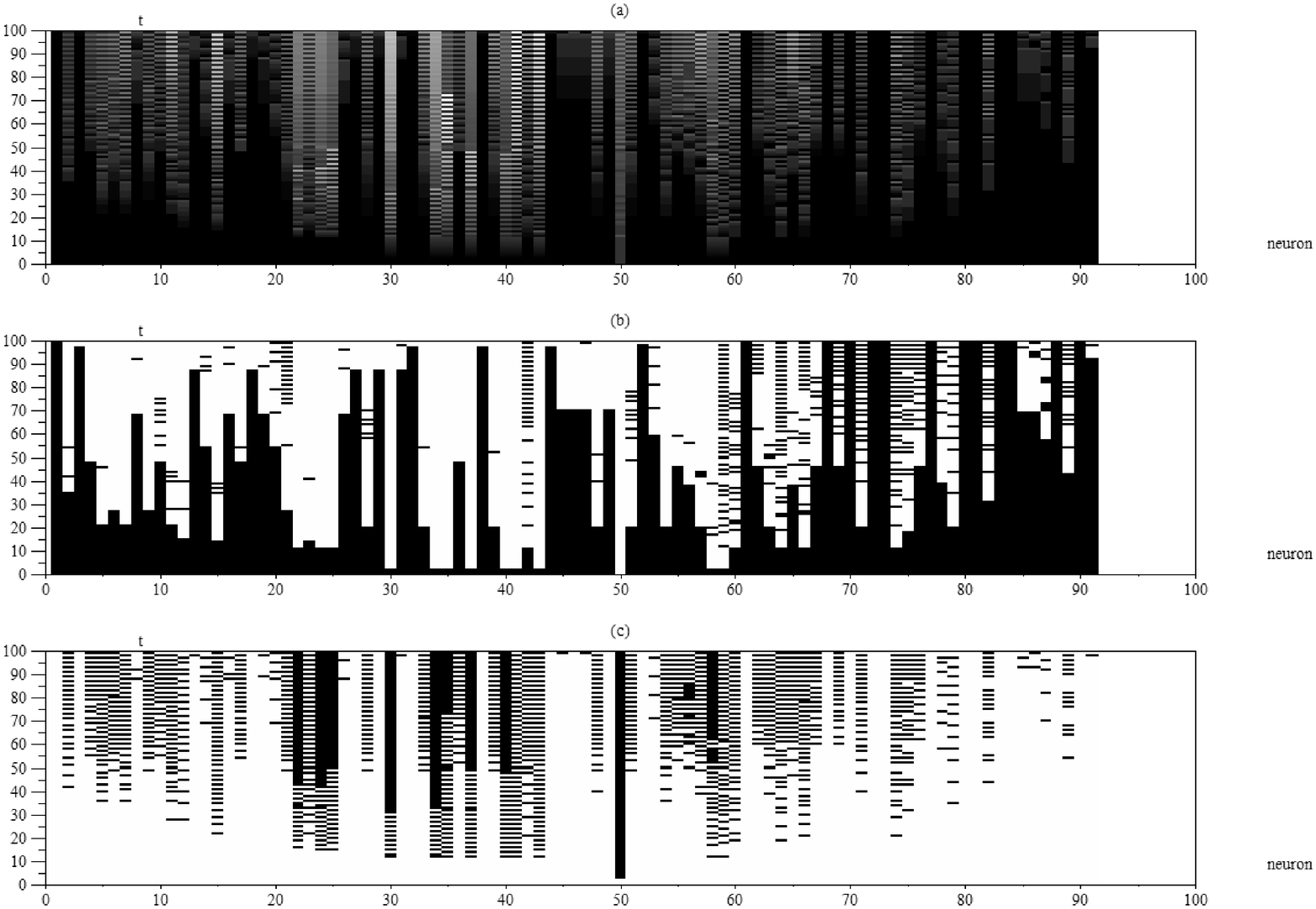}
   \caption{Activation diagrams for GG network: (a) probability
  activation of each node along time (a); the active and non-activate neurons
  along time (b); and the spikes produced by each neuron
  along time (c).  The source was placed at node 50.
  }~\label{fig:grams_GG} \end{center}
\end{figure}

\begin{figure}[htb]
  \vspace{0.3cm} \begin{center}
  \includegraphics[width=1\linewidth]{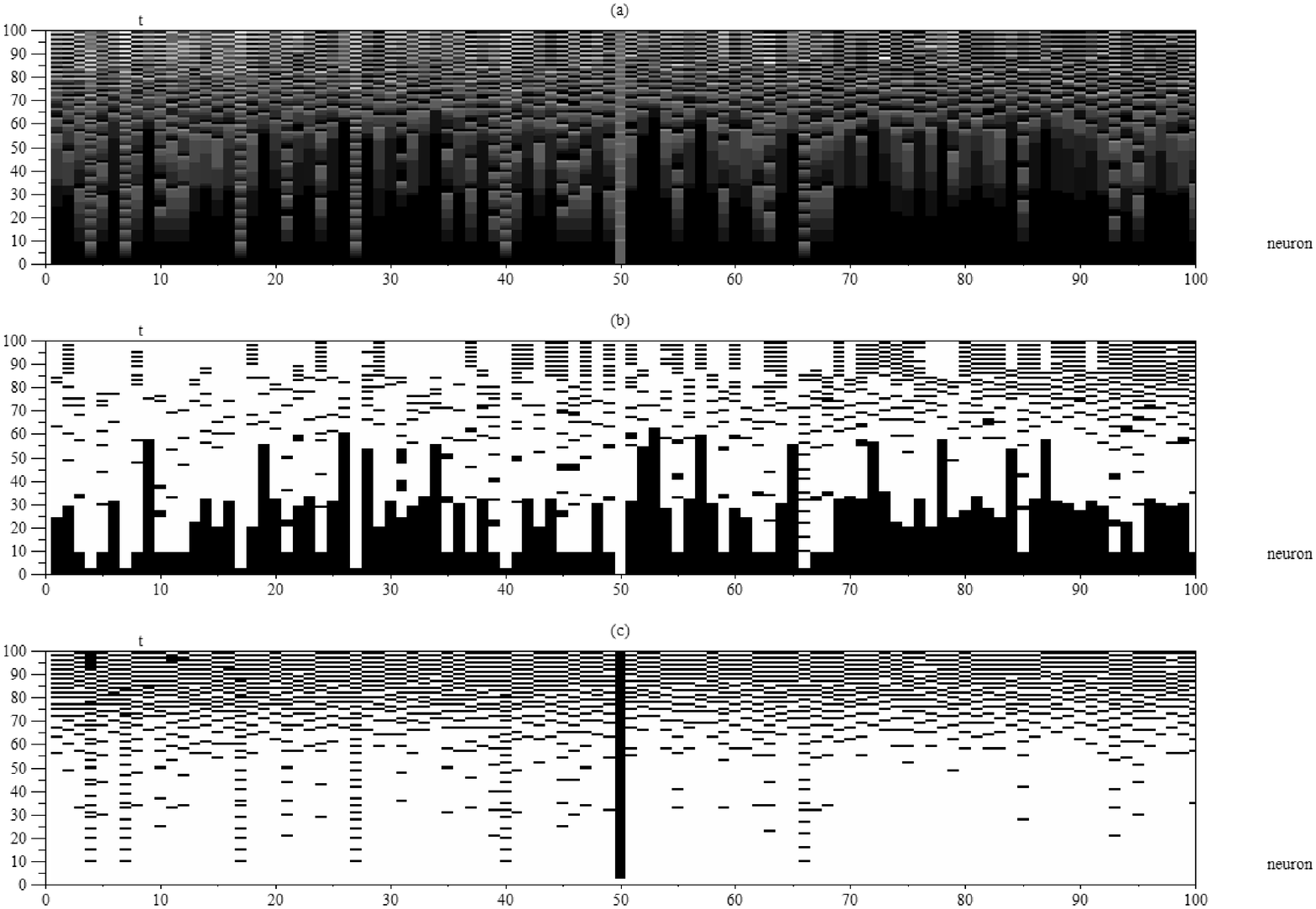}
   \caption{Activation diagrams for PN network: (a) probability
  activation of each node along time (a); the active and non-activate neurons
  along time (b); and the spikes produced by each neuron
  along time (c).  The source was placed at node 50.
  }~\label{fig:grams_PN} \end{center}
\end{figure}

\begin{figure}[htb]
  \vspace{0.3cm} \begin{center}
  \includegraphics[width=1\linewidth]{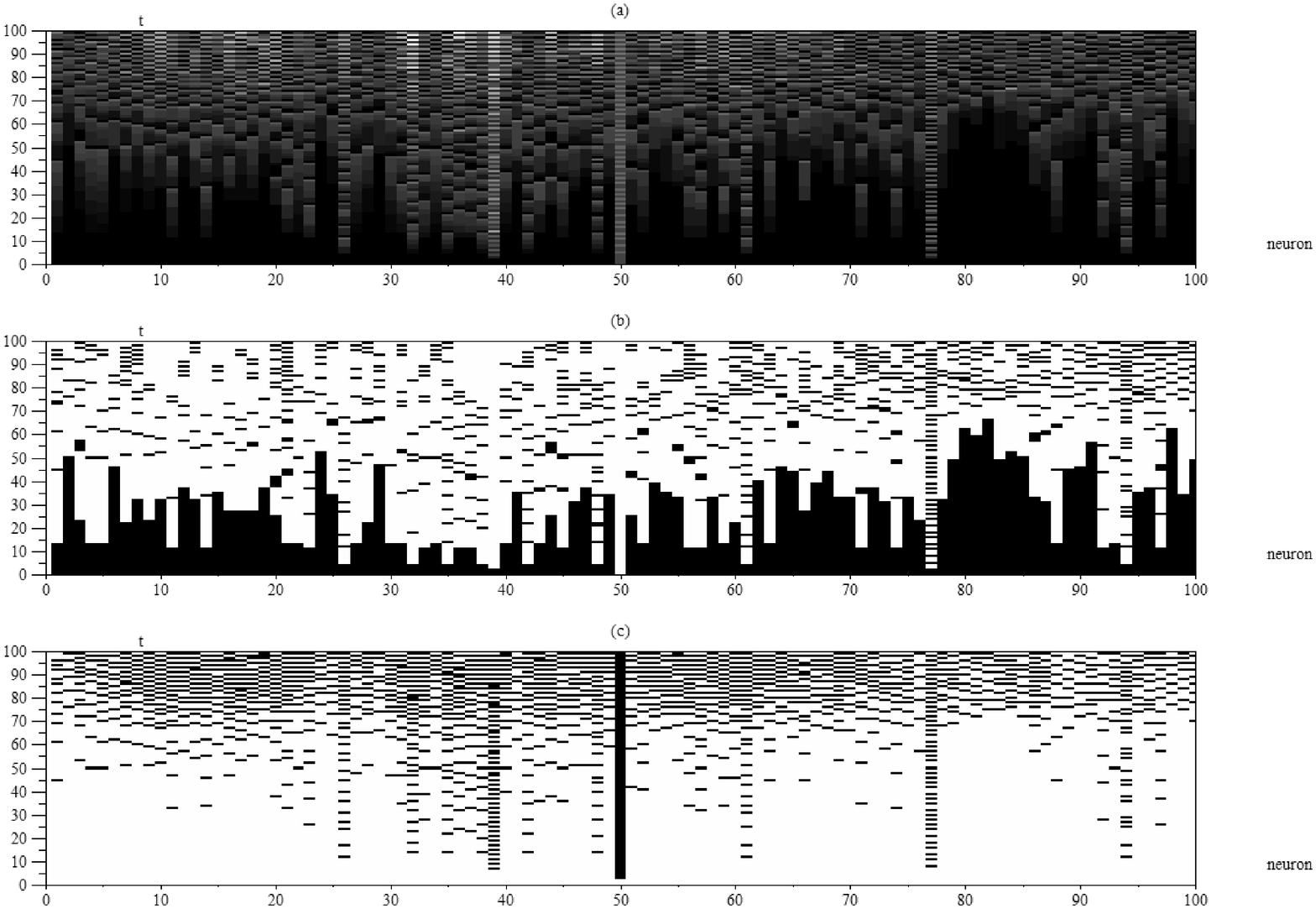}
   \caption{Activation diagrams for PA network: (a) probability
  activation of each node along time (a); the active and non-activate neurons
  along time (b); and the spikes produced by each neuron
  along time (c).  The source was placed at node 50.
  }~\label{fig:grams_PA} \end{center}
\end{figure}

Figures~\ref{fig:grams_ER} to~\ref{fig:grams_PA} show the patterns of
activation of all neurons for each time $t = 1, 2, \ldots, 100$
obtained for each of the 6 networks by having the source of activation
placed at node 50 (with a few exceptions, similar patterns were
identified for the source at other nodes).  More specifically, each of
these figures show the activogram, i.e. the activation at each neuron
along time (a), the active (white) and non-activate (black) neurons
along time; and (c) the spikes produced by each neuron along time.
The activations, instead of the normalized probabilities of
activation, are shown in the activograms in all figures for the sake
of better visualization.

A series of interesting results and interpretations can be identified
from these figures.  As expected, all activations tended to spread
progressively from the source node 50 as time passes, with the rate of
spikes increasing steadily with time.  However, quite distinct patterns
of activation have been observed for each of the considered networks.
In the case of ER (Fig.~\ref{fig:grams_ER}), for instance, a
reasonably uniform distribution of activation along time was obtained,
with most nodes engaging into activity for the first time at similar
instants (after approximately 10 or 20 steps, see
Figure~\ref{fig:grams_ER}(b)).  As shown in Fig.~\ref{fig:grams_BA},
quite a different dynamics of activation has been obtained for the BA
network.  Because of the presence of hubs, several nodes are activated
relatively soon (less than 10 steps), while some nodes are only
recruited much later.  In addition, the activity along time tends to
concentrate in the hubs at the left-hand side of the image.  It would
be particularly interesting to verify whether the rate of individual
activations follow a power law.  Yet another pattern of activations
has been obtained for the WS model (Fig.~\ref{fig:grams_WS}), which
has been made clearer by the fact that the original neurons in the
one-dimensional regular lattice used to derive this network had been
sequentialy numbered.  Now, the activation proceeds gradually through
the successive neighbors.  Also, once activated, the neurons seem to
engage in more regular patterns of spiking than those obtained for ER
and BA.  The onset of activation in the GG network is peculiar,
especially regarding the fact that neurons tend to start activity at
the most diverse times.  This is explained by the fact that the GG is
not small-world, implying the activation to progress along the
adjacencies along the network.  Therefore, neurons which are connected
to the source through longer shortest paths will engage into activity
later. The activation diagrams obtained for the PN and PA networks are
remarkably similar, being characterized by relatively uniform times
for activation onset. This result is particularly surprising because
these two networks are know to have markedly distinct
structures~\cite{Costa_comp:2007, Costa_longest:2007}.

\begin{figure*}[htb]
  \vspace{0.3cm} \begin{center}
  \includegraphics[width=1\linewidth]{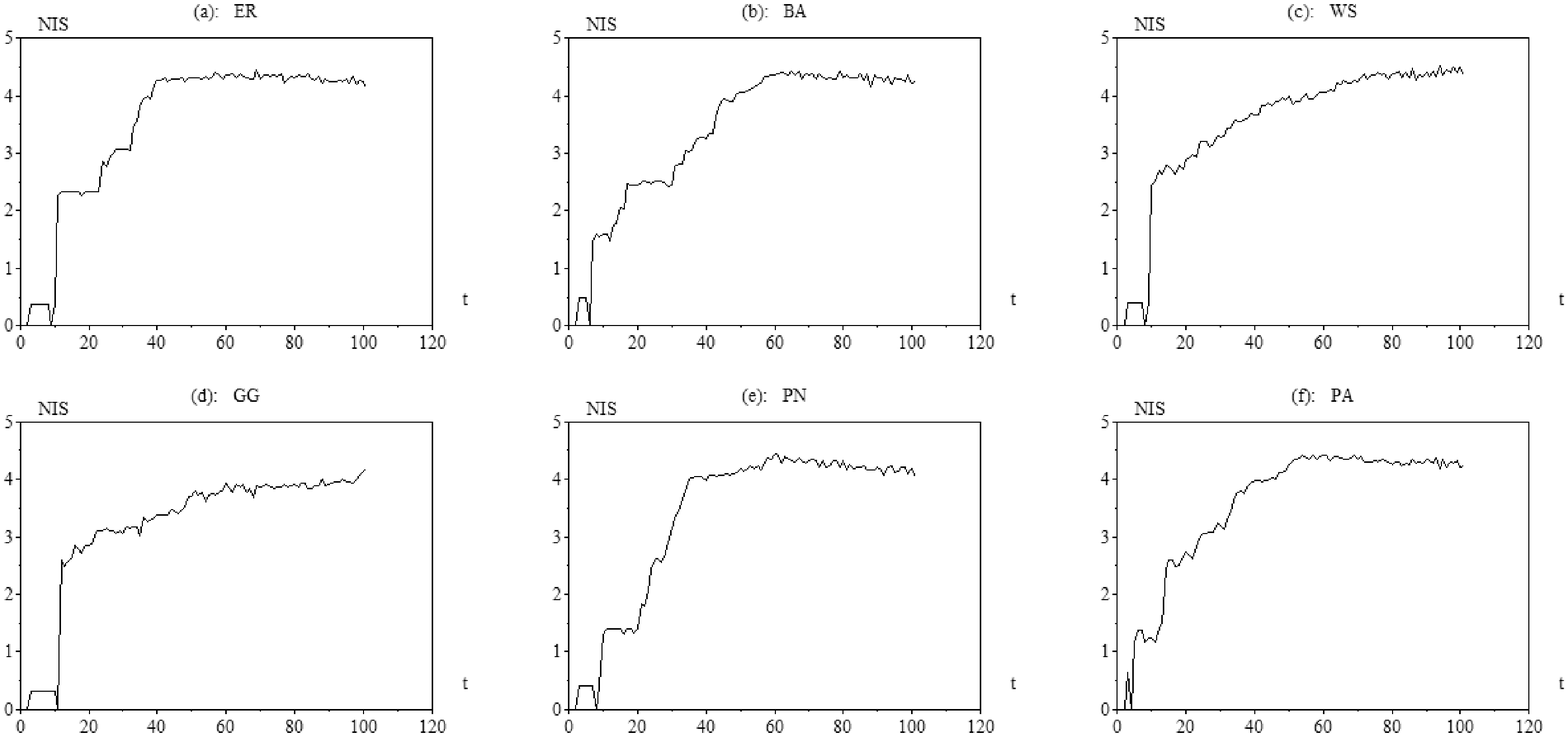}
   \caption{The normalized entropy of the activations along time for each of
       the considered networks.  The source was placed at node 50.
  }~\label{fig:entrs_50} \end{center}
\end{figure*}

\begin{figure*}[htb]
  \vspace{0.3cm} \begin{center}
  \includegraphics[width=1\linewidth]{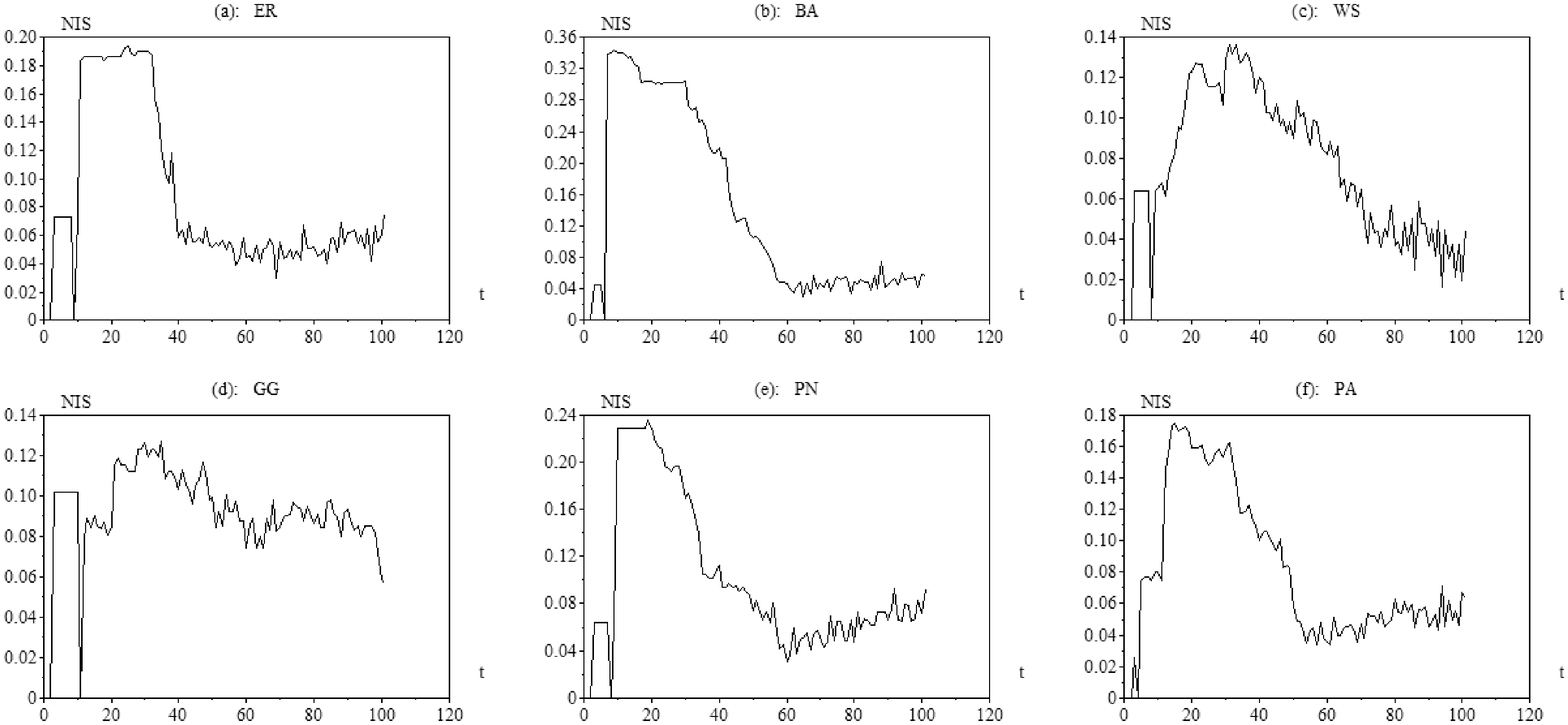}
   \caption{The normalized instantaneous synchronization (NIS) along 
             time for each of the considered networks.  The source was 
             placed at node 50.
  }~\label{fig:NISs_50} \end{center}
\end{figure*}

The normalized entropies of activations for each time considering all
neurons, with source of activity at neuron 50, are shown in
Figure~\ref{fig:entrs_50}.  With a few exceptions, similar patterns
were obtained when the activation source was placed at other neurons.
These curves tend to be similar, involving an initial stage with
plateaux of relatively low entropy, followed by more gradual
progression to higher entropy approaching the maximum limit of $log(N)
\approx 4.61$ in all cases.  Such initial plateaux are mainly a
consequence of the distribution of activities among the axons of the
neurons more immediately connected to the source, especially in the
cases where the node associated to the source, or its more immediate
neighbors, had large out-degree (recall that this implies the activity
to be distributed amongst the outgoing edges).
Figure~\ref{fig:NISs_50} shows the normalized instantaneous
synchronizations (NIS) obtained along time considering all neurons,
with source of activation at node 50.  Interestingly, after exhibiting
some plateaux of synchronization at the initial time steps, these
curves tend to evolve to a peak of synchronization and then decrease
to near zero activation (typically after 60 steps).  Observe the higher
values of NIS obtained for the BA network (Fig.~\ref{fig:NISs_50}b)
and the more gradual decrease of NIS in the GG case
(Fig.~\ref{fig:NISs_50}d).

\begin{figure*}[htb]
  \vspace{0.3cm} \begin{center}
  \includegraphics[width=1\linewidth]{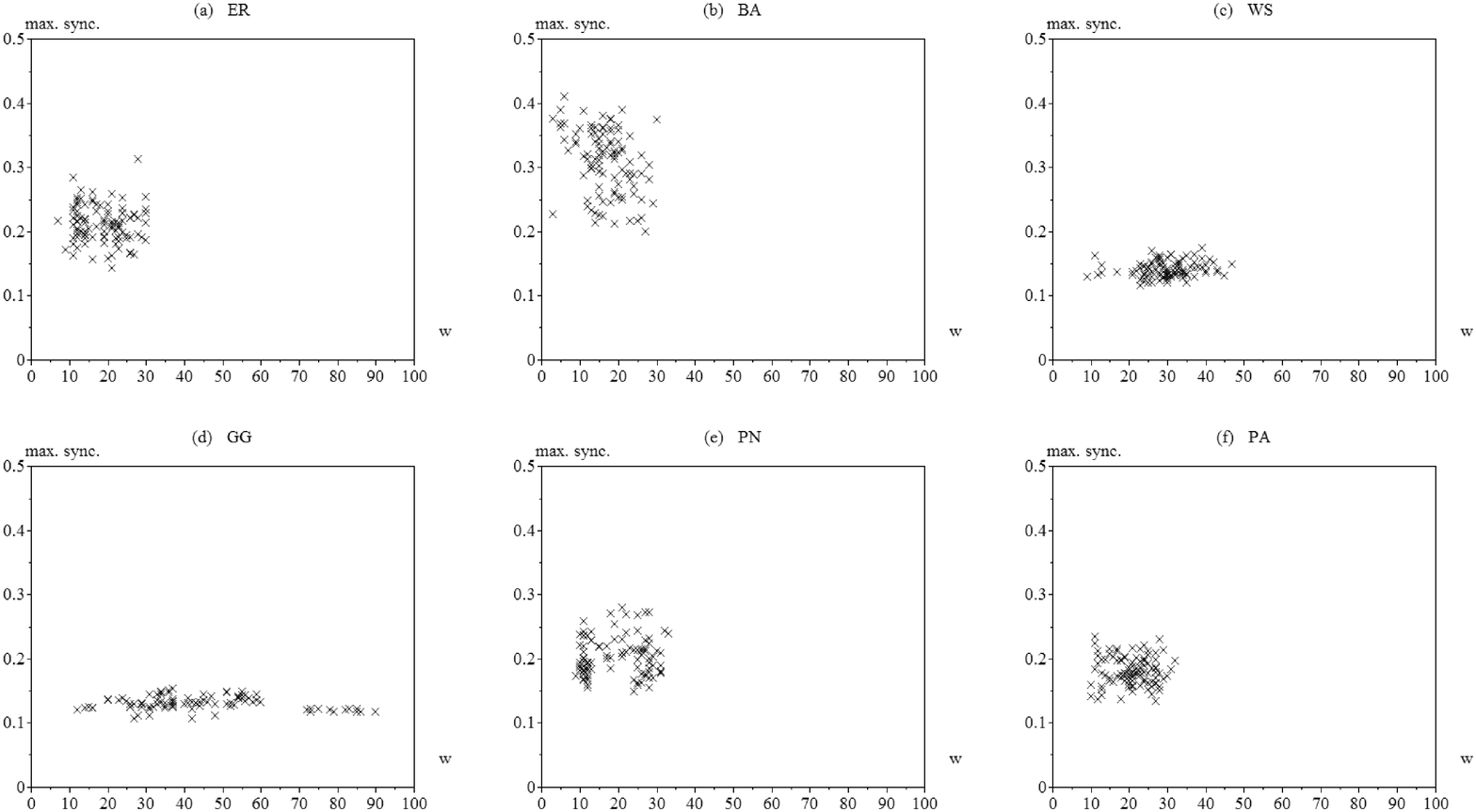}
   \caption{Scatterplos of the maximum NIS ($y-$axes) and time at
              which it manifested itself ($x-$axes) obtained for
              each networks by considering activations at each
              possible node.
  }~\label{fig:max_syncs} \end{center}
\end{figure*}

A more complete picture of the instantaneous synchronization of the
networks can be obtained by inspecting Figure~\ref{fig:max_syncs},
where the $x-$axes correspond to the maximum NIS obtained along all
times, all nodes, and by considering the source in any node; while the
$y-$axes show the time $w$ at which the maximum NIS was obtained.
Remarkably, each of the networks led to relatively homogeneous maximum
synchronization and time at which it occurred, defining reasonably
dense clusters of points in these scatterplots.  The times $w$ at
which the maximum synchronizations were observed are similar in all
cases, except the GG network, and tend to be comprised within the
interval from 1 to 30 steps.  In the case of the GG, these times
extend to 100 steps, reflecting the fact that the GG structure does
not present the small-world.  The highest synchronizations were
observed for the BA, and the lowest for the WS and GG cases. The
curves obtained for the PN and PA cases are similar one another and
resemble the ER counterpart.

\begin{figure}[htb]
  \vspace{0.3cm} \begin{center}
  \includegraphics[width=1\linewidth]{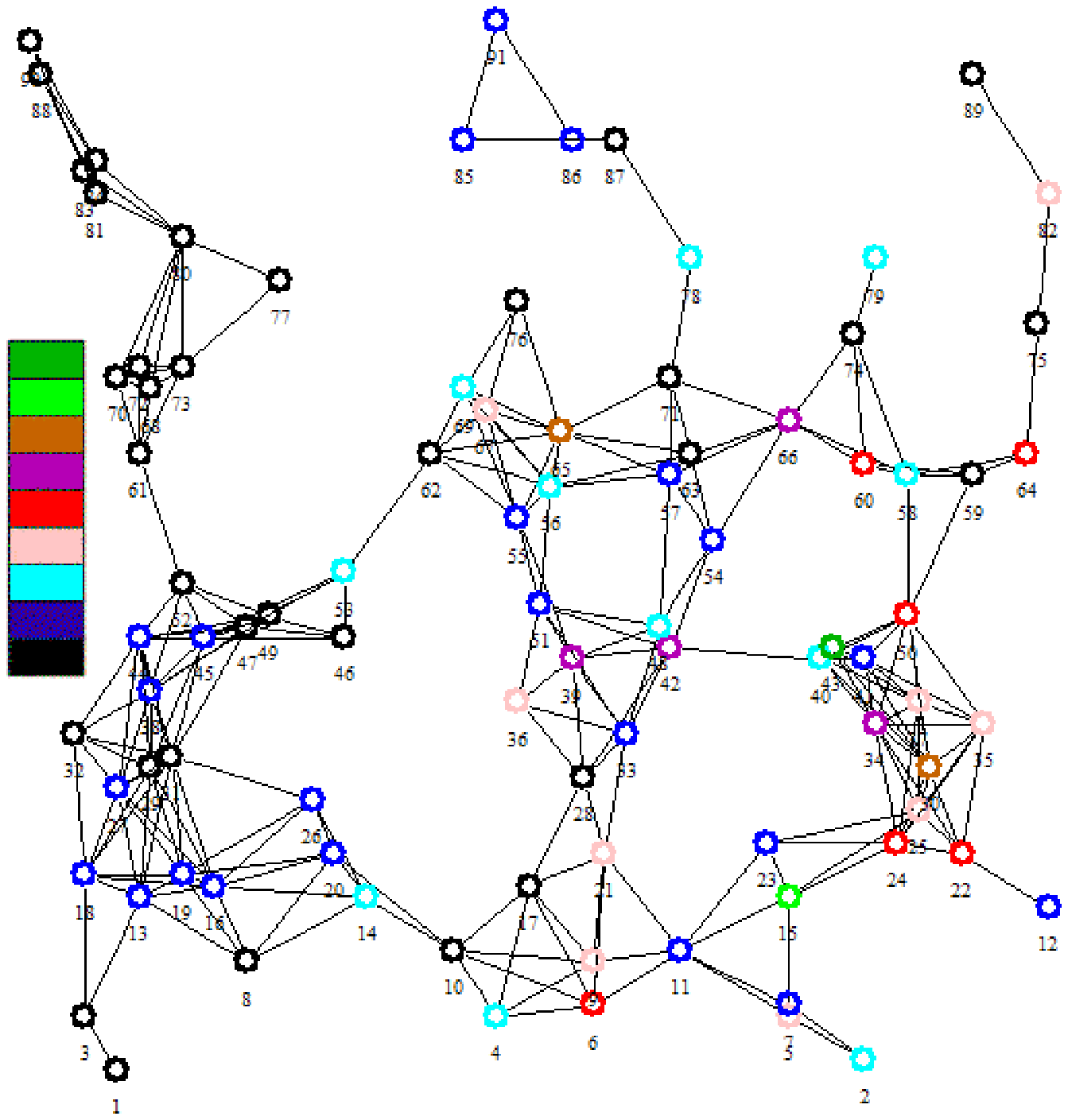} \\
(a) \\ \vspace{0.5cm}
  \includegraphics[width=1\linewidth]{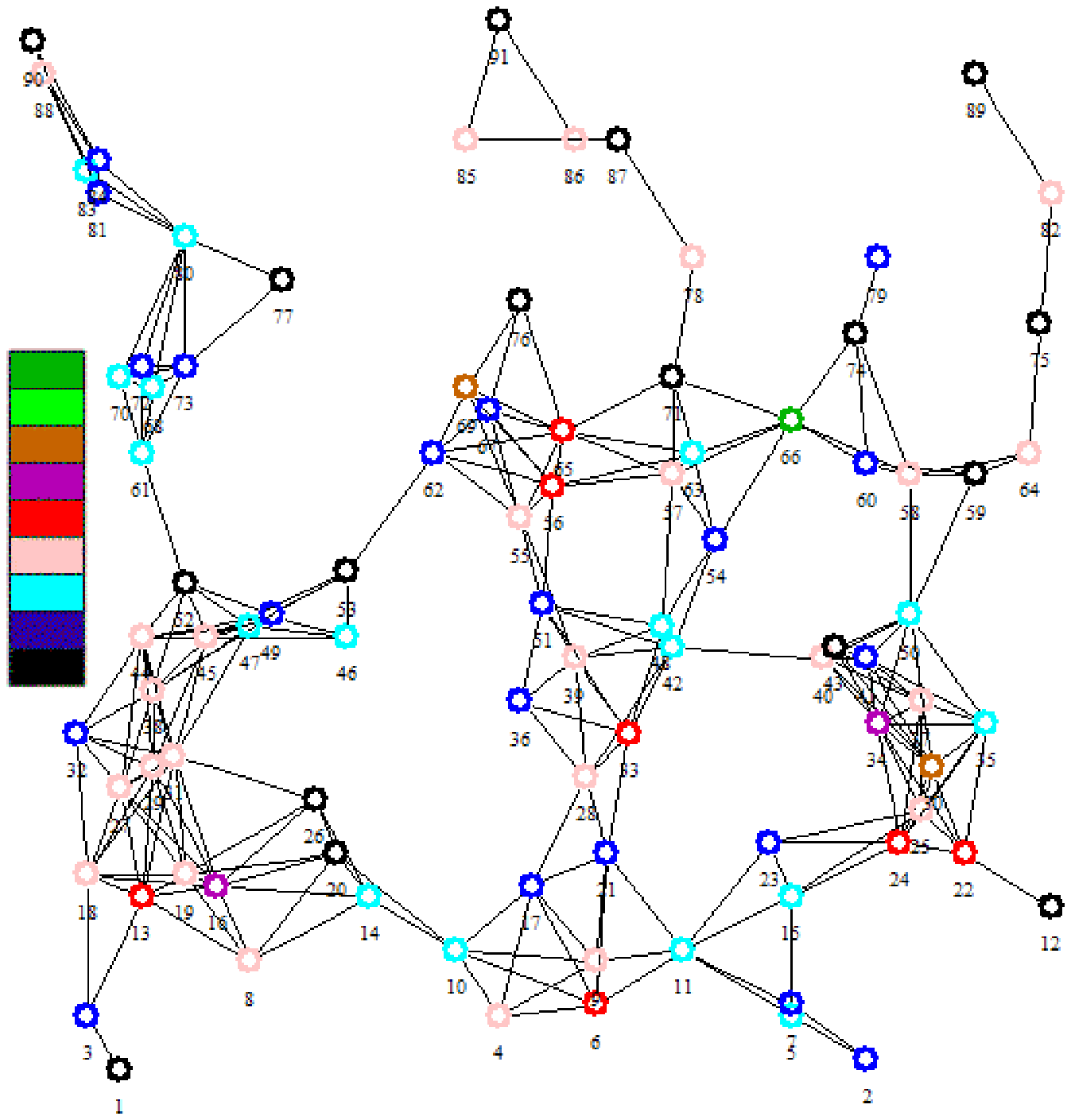}
(b) \\
   \caption{The normalized instantaneous synchronizations at each
               node of the GG network considered in this article
               at times 100 (a) and 1000 (b).
  }~\label{fig:GG} \end{center}
\end{figure}

Figure~\ref{fig:GG} shows the normalized activations $a(i,t)$ at each
of the nodes $i$ of the considered GG network at time 100 (A) and 1000
(b), with the source of activation placed at node 50. Observe that the
pattern of activation around the source tended to be similar in these
two cases, suggesting that the activation tends to a steady-state
which unfolds more strongly around the source.  However, most of the
other activations changed almost completely along the time period
between 100 and 1000 steps.

\begin{figure*}[htb]
  \vspace{0.3cm} \begin{center}
  \includegraphics[width=1\linewidth]{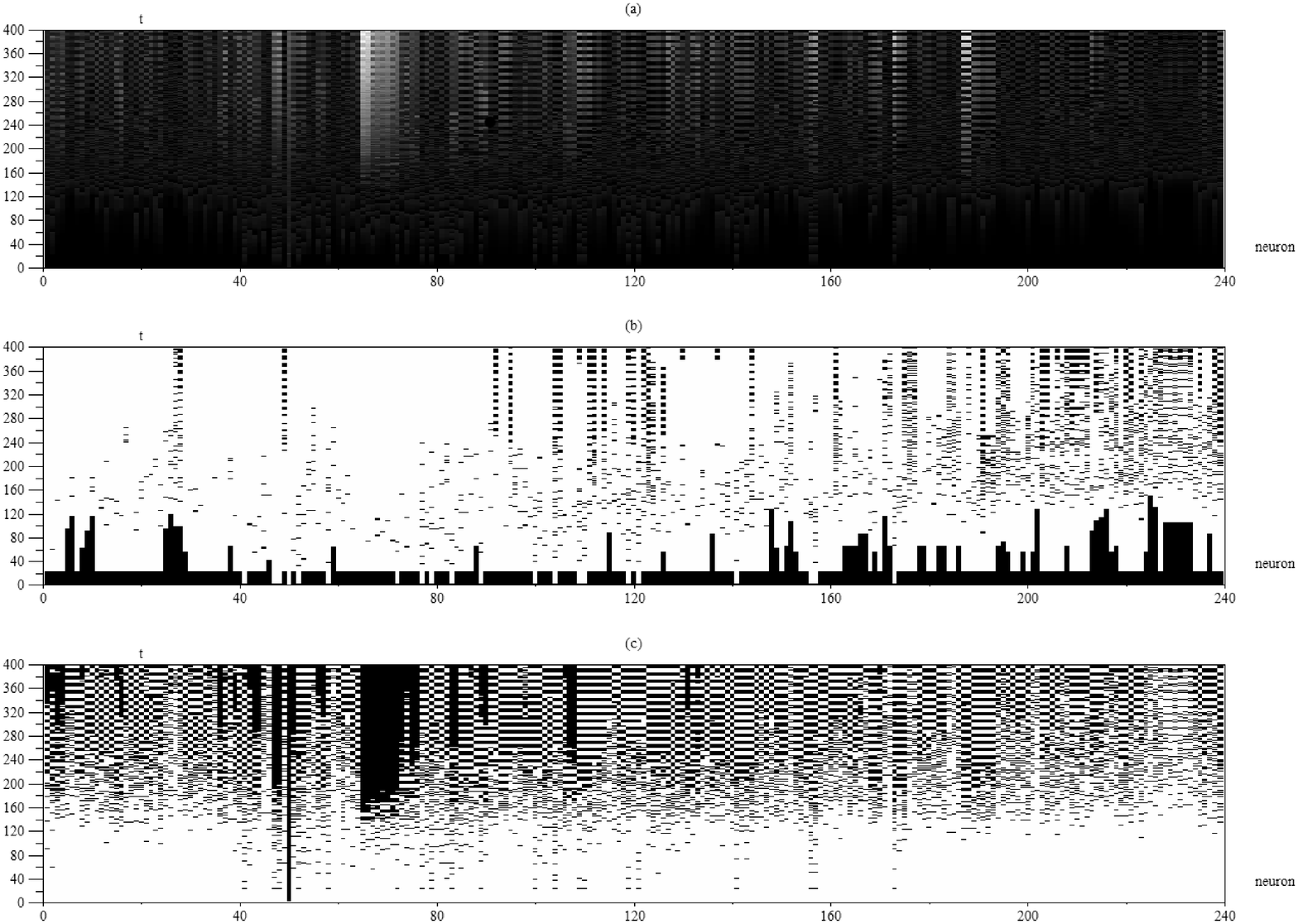}
   \caption{The activation diagrams for the 239 nodes in the largest
            connected component of the \emph{C. elegans} network.
  }~\label{fig:grams_cel} \end{center}
\end{figure*}

Figure~\ref{fig:grams_cel} shows the activation diagrams (a),
activated nodes (b) and spikes (c) obtained for all the 239 neurons of
the \emph{C. elegans} network along the 100 initial time steps, with
the activation source at node 50.  As can be clearly observed from
Figure~\ref{fig:grams_cel}, most of the nodes are engaged in activity
after only about 20 time steps.  This seems to be related to the
relatively high average node out-degree (14.36) of this network.
Interestingly, the activation of the neurons seem to undergo an abrupt
increase after approximately 100 steps, with most neurons presenting
similar frequencies of activation thereafter.  In addition, several
neurons tended to exhibit similar frequency of spikes after such a
transition.  The peak of activation at the latest stages occurs at the
maximum hub of this network (out-degree 73).

\begin{figure}[htb]
  \vspace{0.3cm} 
  \begin{center}
  \includegraphics[width=1\linewidth]{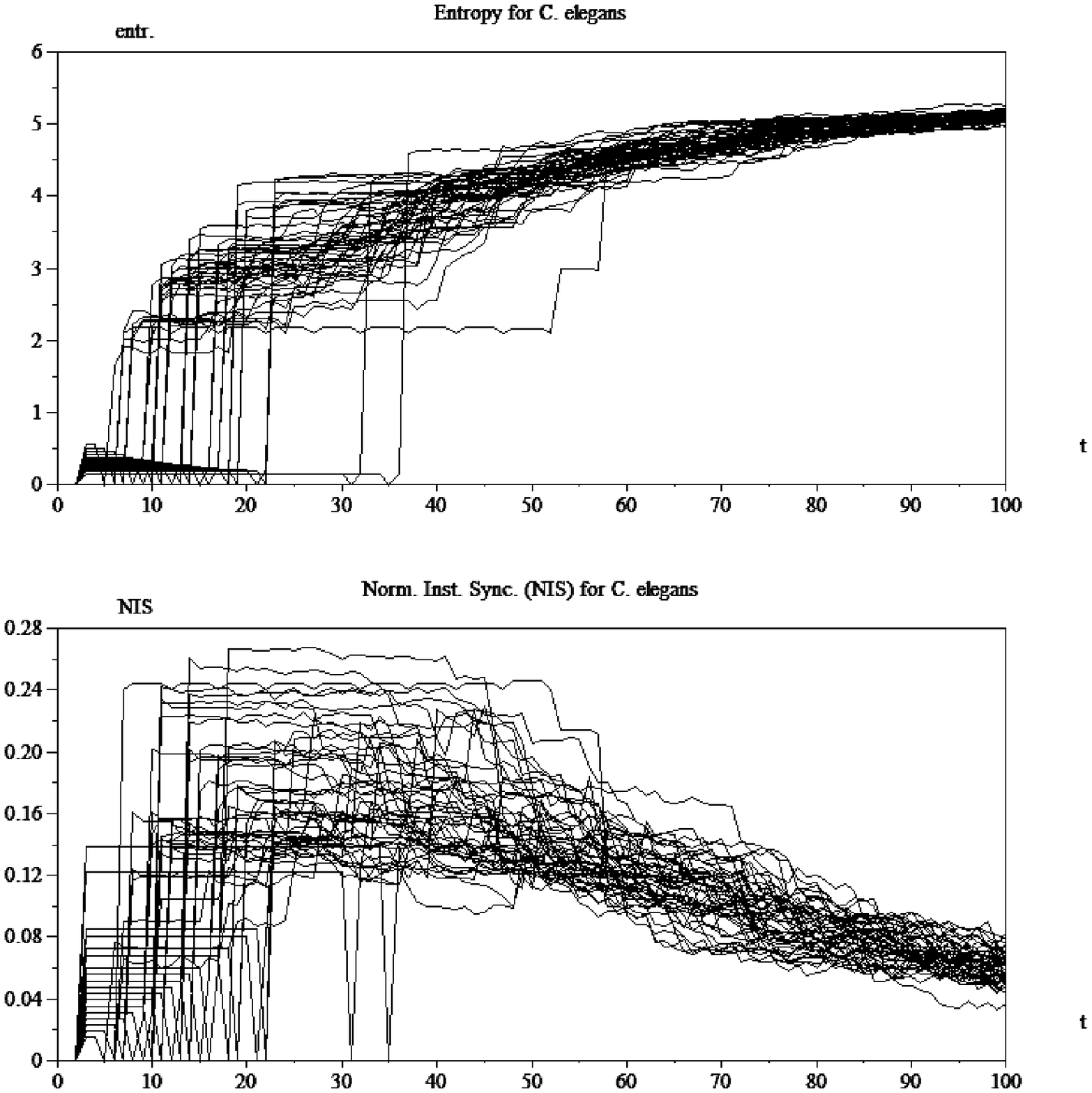} \\
   \caption{The entropies and NISs along time for neurons 1 to 50
                in the \emph{C. elegans} network.
  }~\label{fig:entr_sync_cel} \end{center}
\end{figure}

The evolutions of entropy and NIS along time with activations placed
at nodes 1 to 50, shown in Figure~\ref{fig:entr_sync_cel}, are
particularly diversified, with initial plateaux presenting quite
different lengths and heights.  Interestingly, the lengthier plateaux
also tended to yield the highest normalized instantaneous
synchronization.  Also noticeable are the shorter and lower plateaux
which appear from time stapes 1 to 20.  However, after nearly 60 steps
the synchronizations collapse.

\begin{figure}[htb]
  \vspace{0.3cm} 
  \begin{center}
  \includegraphics[width=1\linewidth]{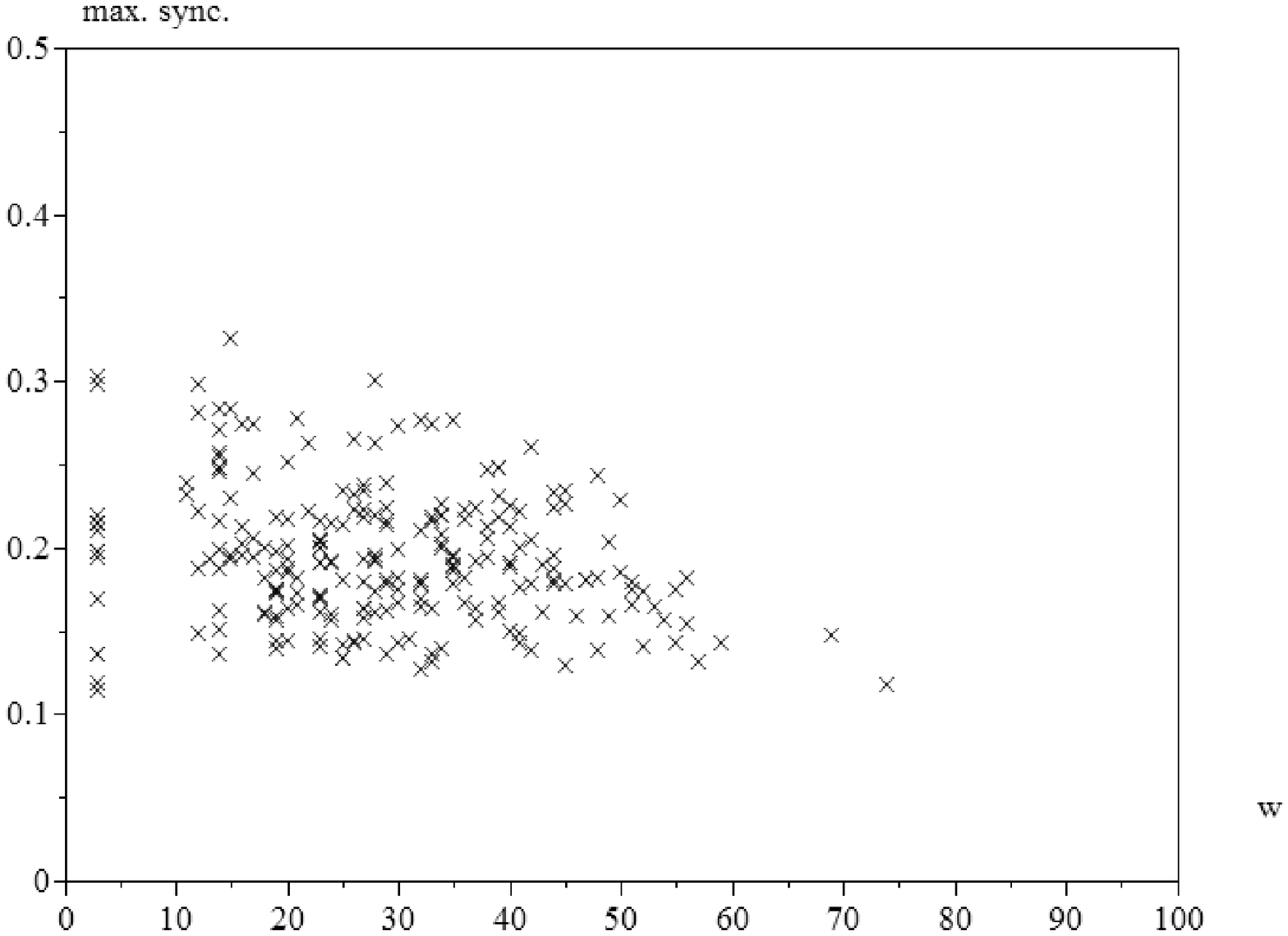} 
  \caption{The
  maximum NIS and respective times obtained along the initial 100
  steps for the \emph{C. elegans} network, considering the source
  placed at each of the nodes.  }~\label{fig:max_cel} 
  \end{center}
\end{figure}

The maximum NISs and respective times at which they occurred for the
activation placed at each of the nodes in the \emph{C. elegans}
network are shown in Figure~\ref{fig:max_cel}.  It can be inferred
from this figure that the \emph{C. elegans} presents moderate values
of maximum normalized instantaneous synchronizations, which occurs at
an intensity comparable to that obtained for the ER, PN and PA
networks.  Interestingly, in some cases the maximum synchronization
took place as late as at the 67th and 73th steps.

\section{Oscillations in Uniformly-Random Networks}

In this section we consider a series of complementary approaches in
order to characterize the transient and equilibrium dynamics in two
types of integrate-and-fire complex neuronal networks --- namely ER
and WS structure.  Representative samples of each of theses types of
networks have been selected with sizes $N = 25, 50$ and $100$ and
average degrees $\left< k \right> = 10, 20, 30, 40$ and $50$.

\begin{figure*}[htb]
  \vspace{0.3cm} 
  \begin{center}
  \includegraphics[width=0.9\linewidth]{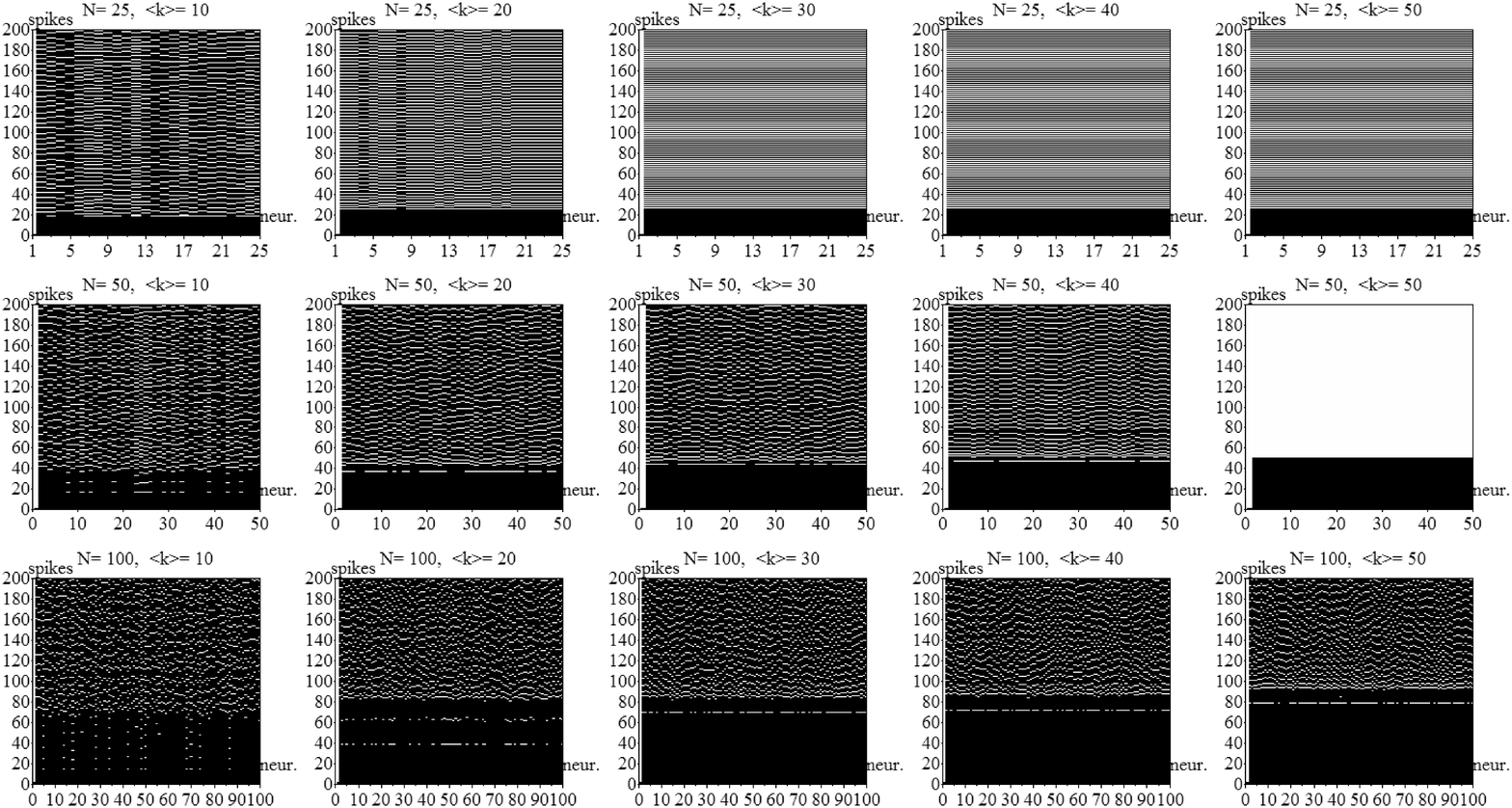} \\  
  \caption{The spikegrams, i.e. the occurrence of spikes for each neuron
  (identified along the $x-$axis) along time ($y-$axis) obtained
  for the ER configurations.  Observe the avalanche transitions during
  the transient regime, as well as the increase of frequency
  regularity and synchronization with observed for larger values of
  average degree.
  }~\label{fig:spikes_ER} 
  \end{center}
\end{figure*}

Figure~\ref{fig:spikes_ER} shows the spikegrams obtained for the
several configurations of ER networks.  Observe the sell-defined
avalanche transitions at the beginning of the spikes obtained for all
network configurations.  In agreement with previous
results~\cite{Costa_equiv:2008}, the avalanches initiation times were
found to be quite similar for all network sizes, irrespectively of the
average degree.  The early spikes obtained for $N=50$ and $\left< k
\right> = 10$; $N=100$ and $\left< k
\right> = 10$; $N=100$ and $\left< k \right> = 20$ and $N=100$;
$\left< k \right> = 30$ and $\left< k \right> = 50$ were a consequence
of the existence of communities which arise in the ER networks arising
from random fluctuations.  Interestingly, the oscillations tended to
become more regular (more similar frequencies) and synchronized with
the increase of the average degree.  The routes to regularity and
synchronization this parameter is increased exhibited groups of
neurons producing similar spiking patterns.  

\begin{figure*}[htb]
  \vspace{0.3cm} 
  \begin{center}
  \includegraphics[width=0.9\linewidth]{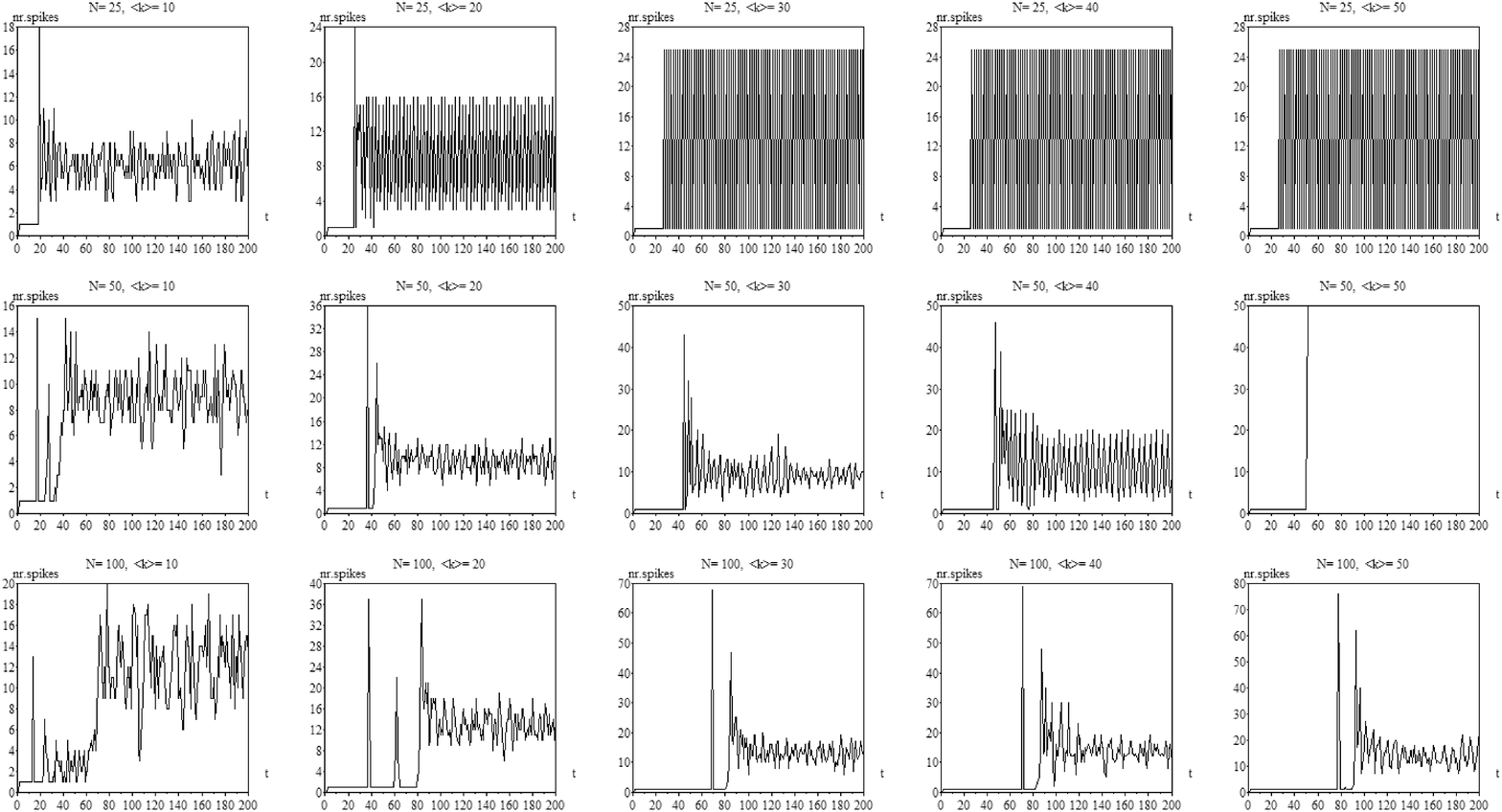} \\  
  \caption{The number of spikes along time obtained for the original
             ER complex network configurations.
  }~\label{fig:nr_spikes_ER} 
  \end{center}
\end{figure*}

The total number of spikes, shown in Figure~\ref{fig:nr_spikes_ER}
reflects the ensemble behavior of the respective complex neuronal
networks, arising as a consequence of the linear superposition of the
spikes being produced by each neuron (analogous to EEG potentials).
The avalanche transitions can be clearly identified, taking place
after nearly 40 steps for $N=25$, 40 steps for $N=50$ and 80 steps for
$N=100$, which confirms the previous study reported
in~\cite{Costa_equiv:2008}.  In several cases, a clear intense peak is
observed at the avalanche time, which is followed by
oscillations whose regularity tend to increase with the average
degree.

\begin{figure*}[htb]
  \vspace{0.3cm} 
  \begin{center}
  \includegraphics[width=0.9\linewidth]{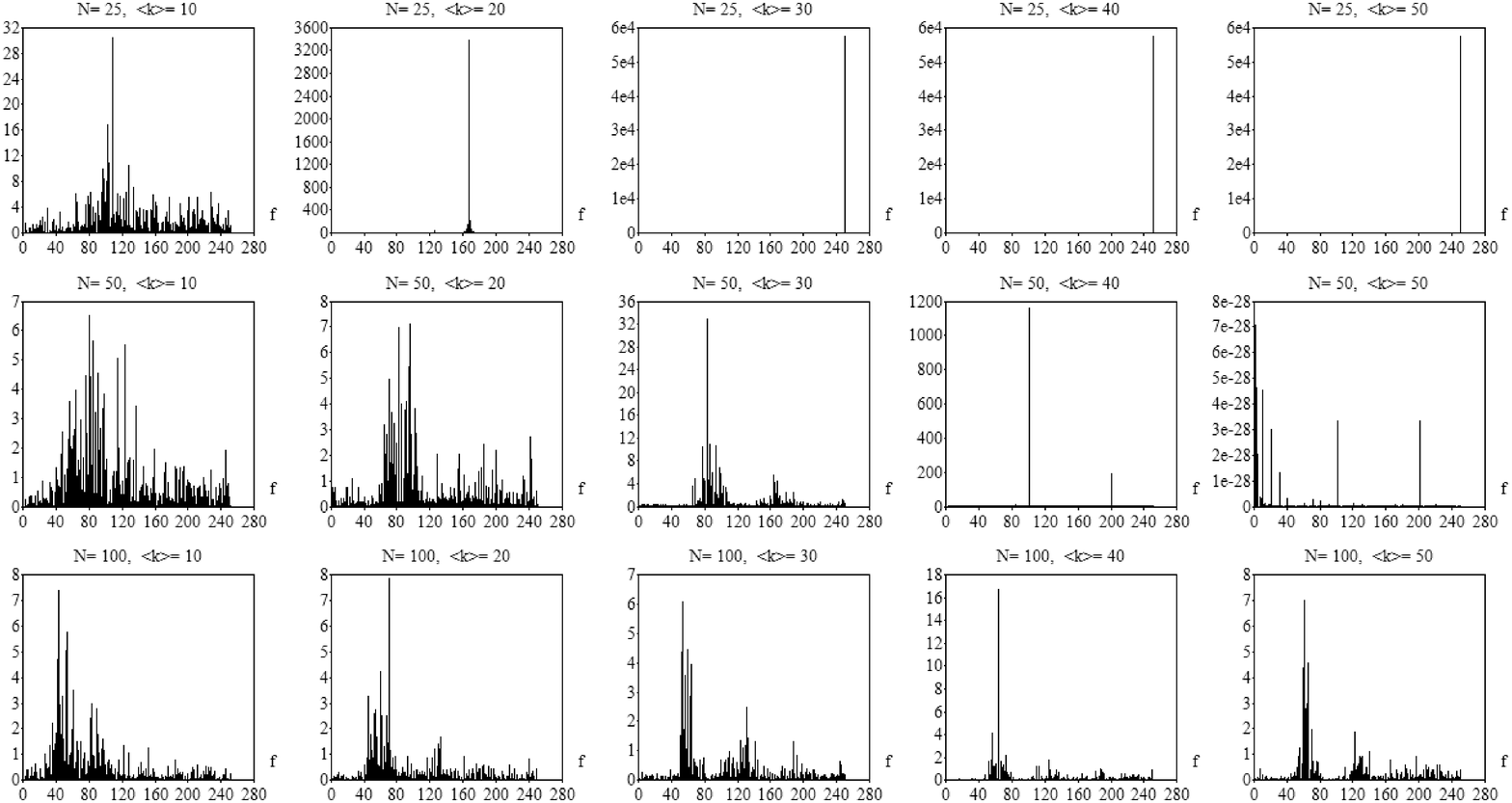} \\ \caption{The
  power spectra obtained for the number of spikes produced by the
  whole original ER networks.  }~\label{fig:spectra_ER} 
  \end{center}
\end{figure*}

In addition to the total number of spikes, it is also important to
take into account the respective power spectra, which are shown in
Figure~\ref{fig:spectra_ER}.  Only the second half of each signal
(number os spikes) has been considered for the calculation of the
power spectra in order to represent the equilibrium regime.  So, a
total $H=500$ time steps have been considered for the estimation of
each spectrum. 

Relatively rich spectral composition can be observed for most cases,
suggesting particularly complex behavior of the oscillatory
components.  The progressive elimination of frequencies with the
increase of the average degree is evident from
Figure~\ref{fig:spectra_ER}, especially for $N=50$ and $N=100$.  In
addition, the peaks of the spectra for $N=50$ and $N=100$ tend to
appear near 20 and 10 frequency units, respectively.  At the same
time, it should be observed that the peaks of the spectra in
Figure~\ref{fig:spectra_ER} tend to shift to the right-hand side for
larger values of the average degree, signaling the increase of the
main frequencies.  The more regular oscillations obtained for larger
average degrees are related to the fact that the higher this
parameter, the more regular the degrees of the networks become.  At
the limit, for very large average degree, all nodes become connected,
implying a highly regular structure as far as all possible topological
features are concerned.

\begin{figure*}[htb]
  \vspace{0.3cm} 
  \begin{center}
  \includegraphics[width=0.9\linewidth]{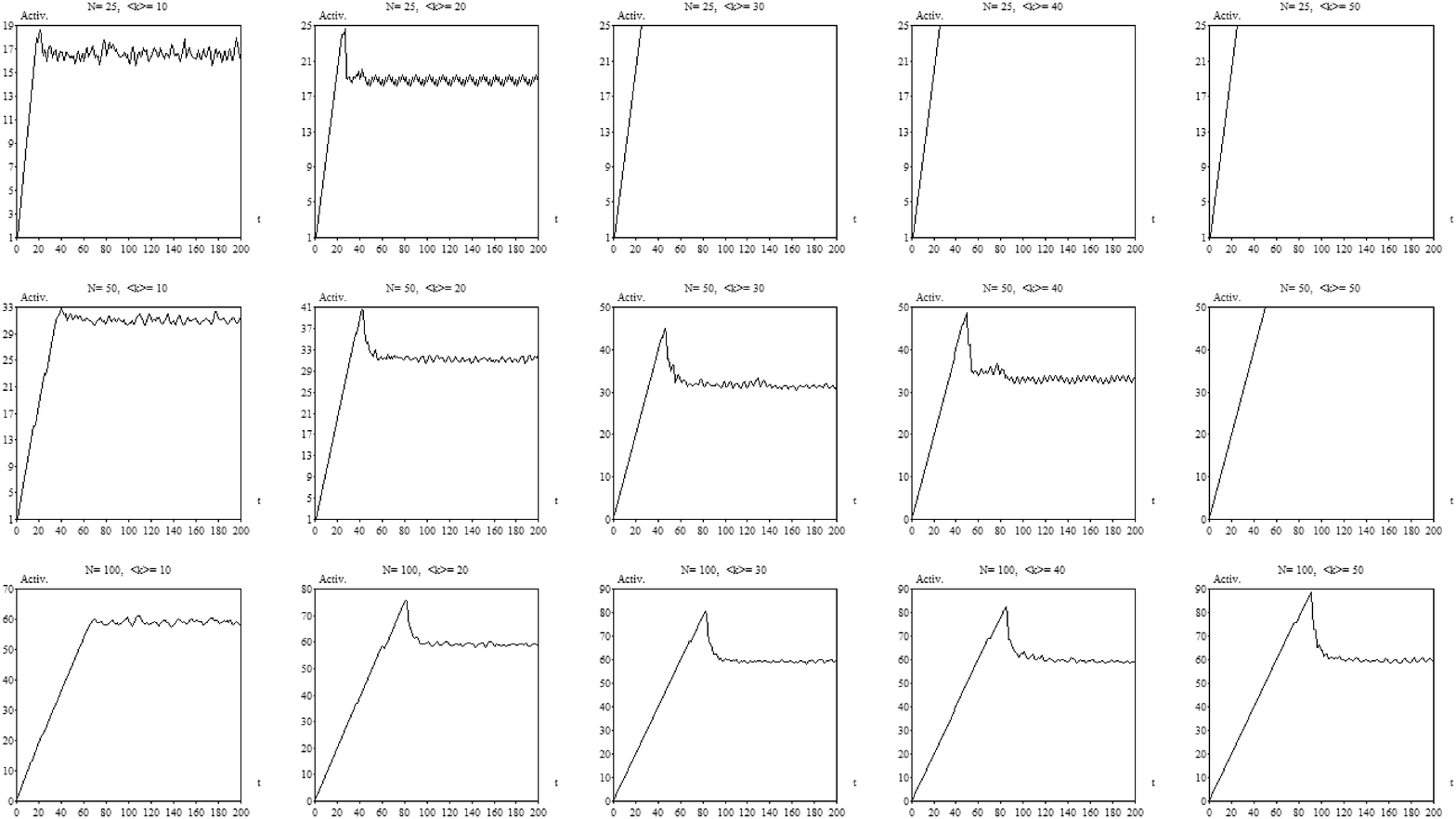} \\  
  \caption{The total activation inside the complex neuronal networks 
                is not conserved (recall that activation is being
                continuously fed from the source node) after the
                main avalanche, but rather
                reaches a plateau of activation after the transient
                regime.
  }~\label{fig:activ_ER} 
  \end{center}
\end{figure*}

Unlike in the previous works~\cite{Costa_nrn:2008, Costa_begin:2008,
Costa_activ:2008, Costa_equiv:2008, Costa_eqcomm:2008}, the activation
inside each neuron is limited to the respective maximum value $L(i)$.
In addition to being more biologically-realistic, such a choice also
allowed particularly interesting oscillatory dynamics.  At the same
time, the limitation of the internal activation implies that the
overall activation, constantly received from the source node, is no
longer guaranteed to be conserved.  It is therefore interesting to
consider the total activation inside the whole complex neuronal
networks along time.  Such a measurement is depicted in
Figure~\ref{fig:activ_ER}.  Interestingly, two clearly distinct
regimes are immediately identified: one transient period in which the
internal activation increases linearly, followed by the steady-state
regime characterized by nearly constant activation (even though
activation is continuously pumped into the system through the source
node). The transition between these two regimes is in most cases
characterized by one abrupt surge of activation, related to the main
avalanche.  Such a result suggests that the main avalanche represents
watershed \emph{signaling the transition of the dynamics in the system
from conservative to dissipative}.

\begin{figure*}[htb]
  \vspace{0.3cm} 
  \begin{center}
  \includegraphics[width=0.9\linewidth]{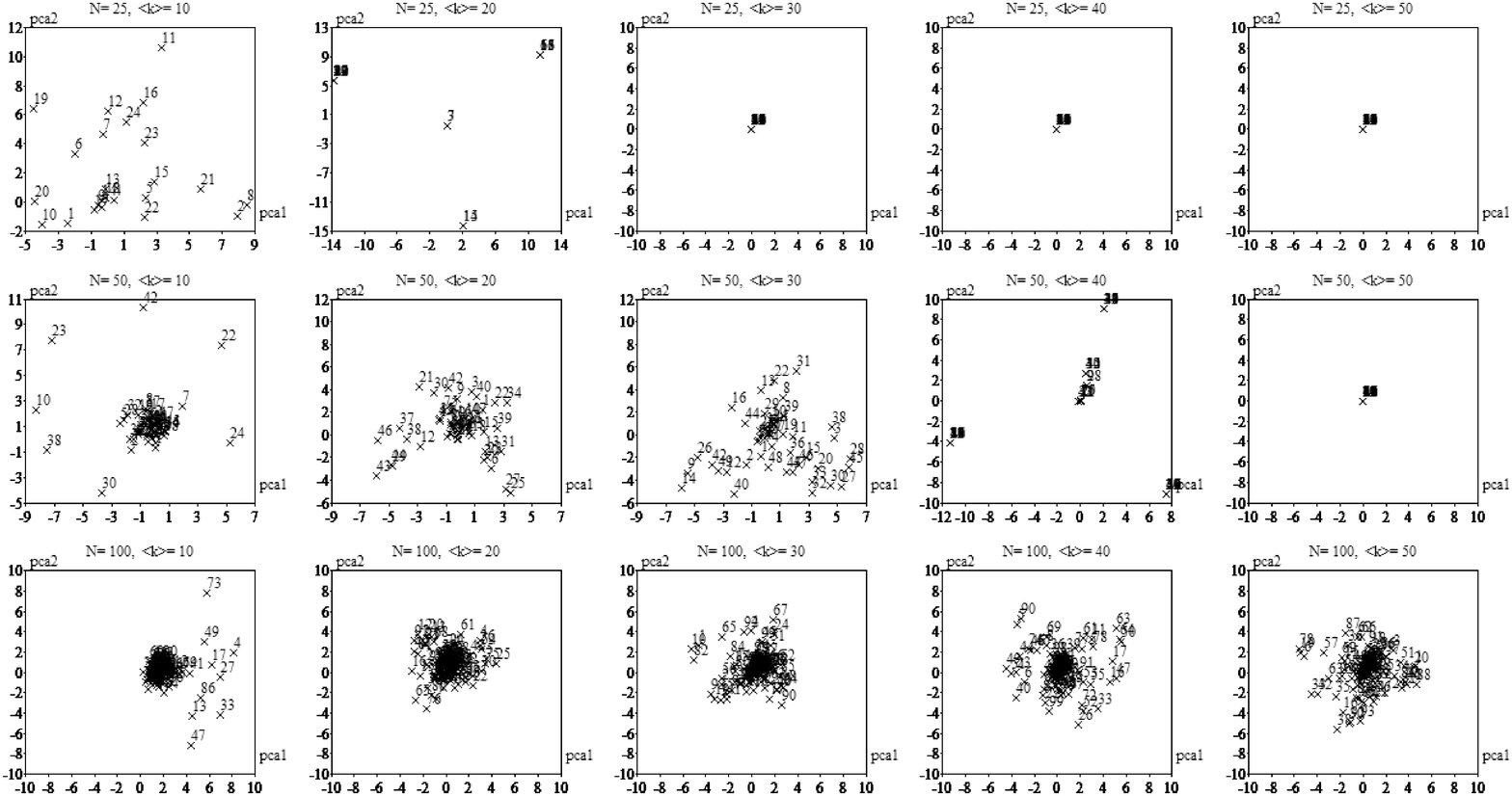} \\  
  \caption{The two-dimensional scatterplots obtained by the application
             of the PCA method over the spike patterns of each neuron
             in the ER configurations.
  }~\label{fig:PCA_ER} 
  \end{center}
\end{figure*}

Because the train of spikes produced by each individual neuron, shown
in Figure~\ref{fig:spikes_ER}, seems to be organized in clusters
during the route towards more regular and synchronous firing
(i.e. similar trains of spikes are obtained for groups of neurons), it
is interesting to investigate such a possibility further.  We do this
with the help of the PCA method, which produced the results shown in
Figure~\ref{fig:PCA_ER}.

It is clear from this figure that the increase of the average node
degree tends to produce denser groups of spiking patterns.  At the
same time, interesting clustering structures are obtained for small
average node degrees, incorporating a dense central cluster (e.g. the
PCA scatterplots for $N= 25$ and $\left< k \right> = 10$, $N= 50$ and
$\left< k \right> = 10$, and $N= 100$ and $\left< k \right> = 10$)
surrounded by outliers.  The dense cluster of spiking patterns tend to
survive even for relatively large average degrees in the case of
$N=100$.

\begin{figure*}[htb]
  \vspace{0.3cm} 
  \begin{center}
  \includegraphics[width=0.9\linewidth]{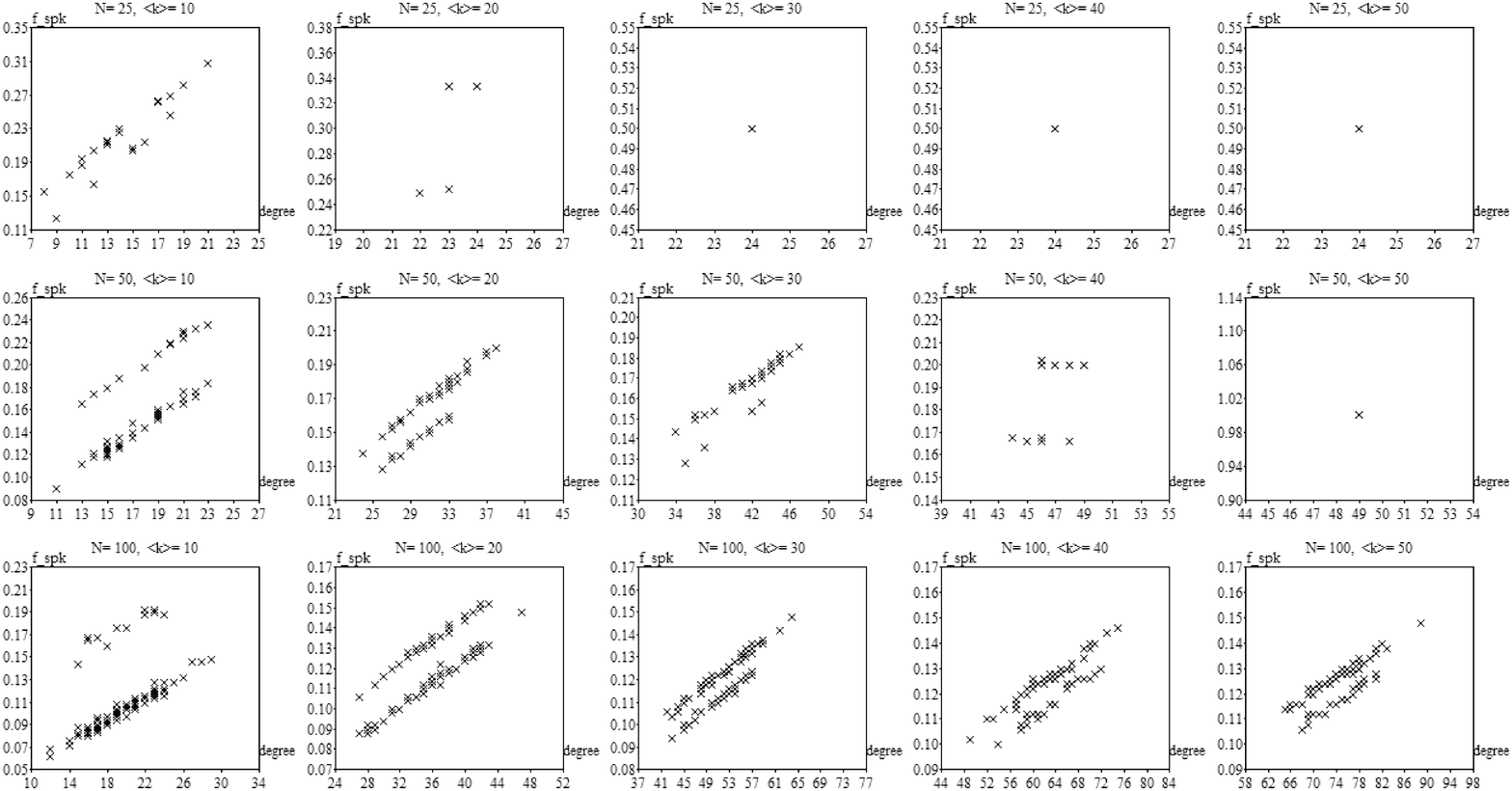} \\  
  \caption{The relationships between the total number of spikes per
             time step
             generated by each neuron and the respective degrees for
             all the considered ER configurations. The number of
             spikes were counted during the last 500 steps of 1000
             steps-long signals.
  }~\label{fig:corrs_ER} 
  \end{center}
\end{figure*}

The correlations between the total number of spikes produced by each
neuron during the simulations and the respective degrees have also
been considered, yielding some of the most interesting results
reported in the present work.  These scatterplots are shown in
Figure~\ref{fig:corrs_ER}.  Striking twin-linear correlation patterns
have been obtained throughout.  Such scatterplots indicate that a node
with a given degree can spike at two distinct frequencies of spikes,
with the number of spikes being linearly related to the node degrees.
At the same time, given one of such rules, the number of spikes (i.e
frequency) tends to be correlated with the degree of the original
nodes.  Such twin correlations are likely to be related to clusters of
original nodes appearing at different concentric hierarchical levels,
as discussed further in Section~\ref{sec:discuss}.

\section{Oscillations in Small-World Networks}

\begin{figure*}[htb]
  \vspace{0.3cm} 
  \begin{center}
  \includegraphics[width=0.9\linewidth]{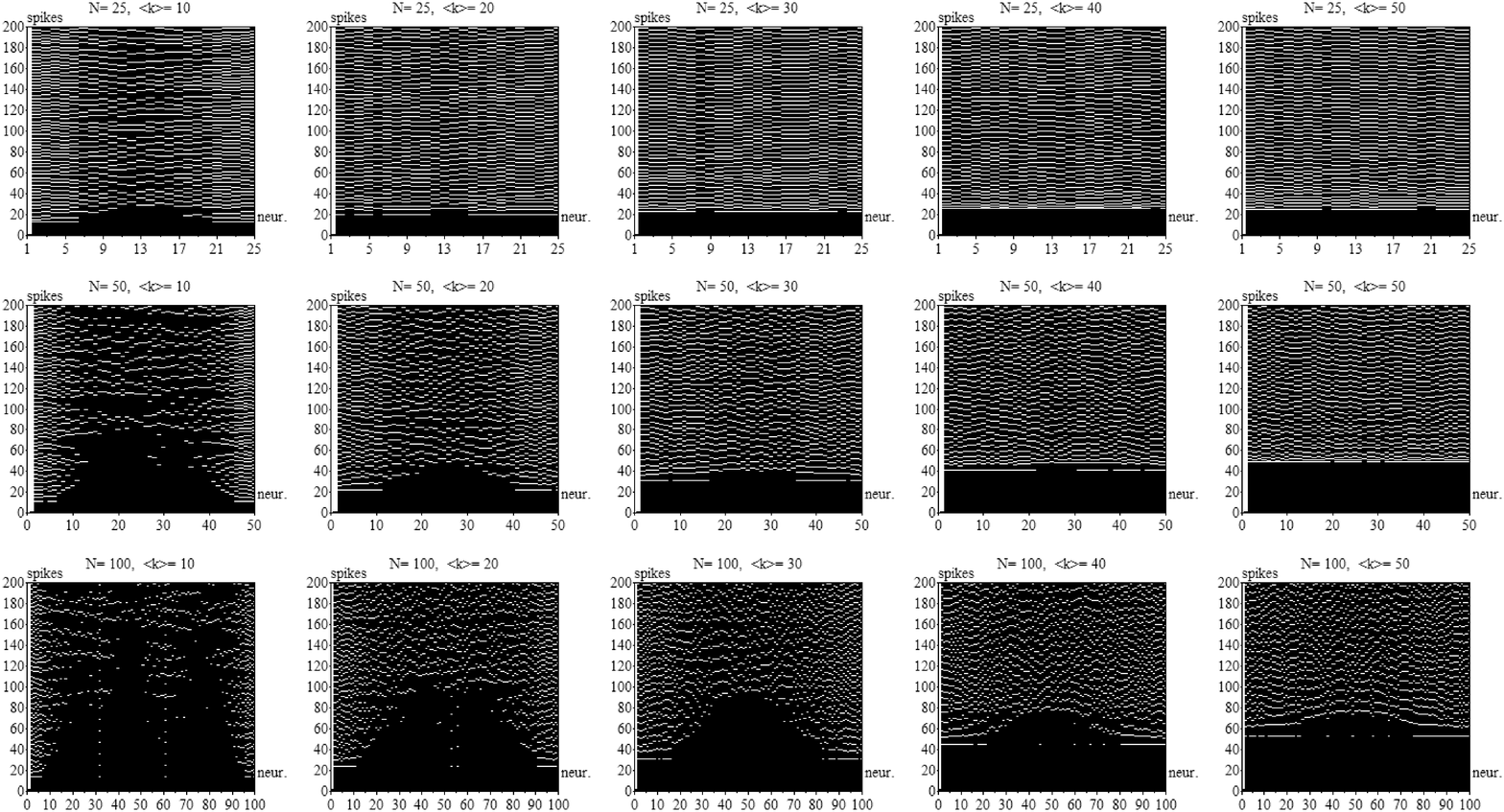} \\  
  \caption{The spikegrams obtained for the WS configurations.
  }~\label{fig:spikes_WS} 
  \end{center}
\end{figure*}

Having obtained a comprehensive characterization of the oscillations
induced along the equilibrium regime in ER networks, it is interesting
to shift our attention to the steady-state dynamics of WS
integrate-and-fire activations.  Figure~\ref{fig:spikes_WS} shows the
spikegrams obtained for the several parametric configurations of the
WS models.  

Several interesting features are evident from the results shown in
Figure~\ref{fig:spikes_WS}.  Foremost are the more gradual avalanches,
which tend to be centralized around the source node (node
1)~\footnote{Though the order of the neurons in such diagrams is
arbitrary, the way in which the WS networks were generated in this
work tends to yield respective adjacency.}. Longer avalanche
initiation times are implied by larger network sizes $N$.  Because the
increase of the average degree implies more degree-regular networks,
the gradual activation of the neurons observed for relatively smaller
average node degrees became sharper for larger average node degrees.
As with the ER cases, the spiking patterns also tended to become more
regular and synchronized for larger values of average degree.

\begin{figure*}[htb]
  \vspace{0.3cm} 
  \begin{center}
  \includegraphics[width=0.9\linewidth]{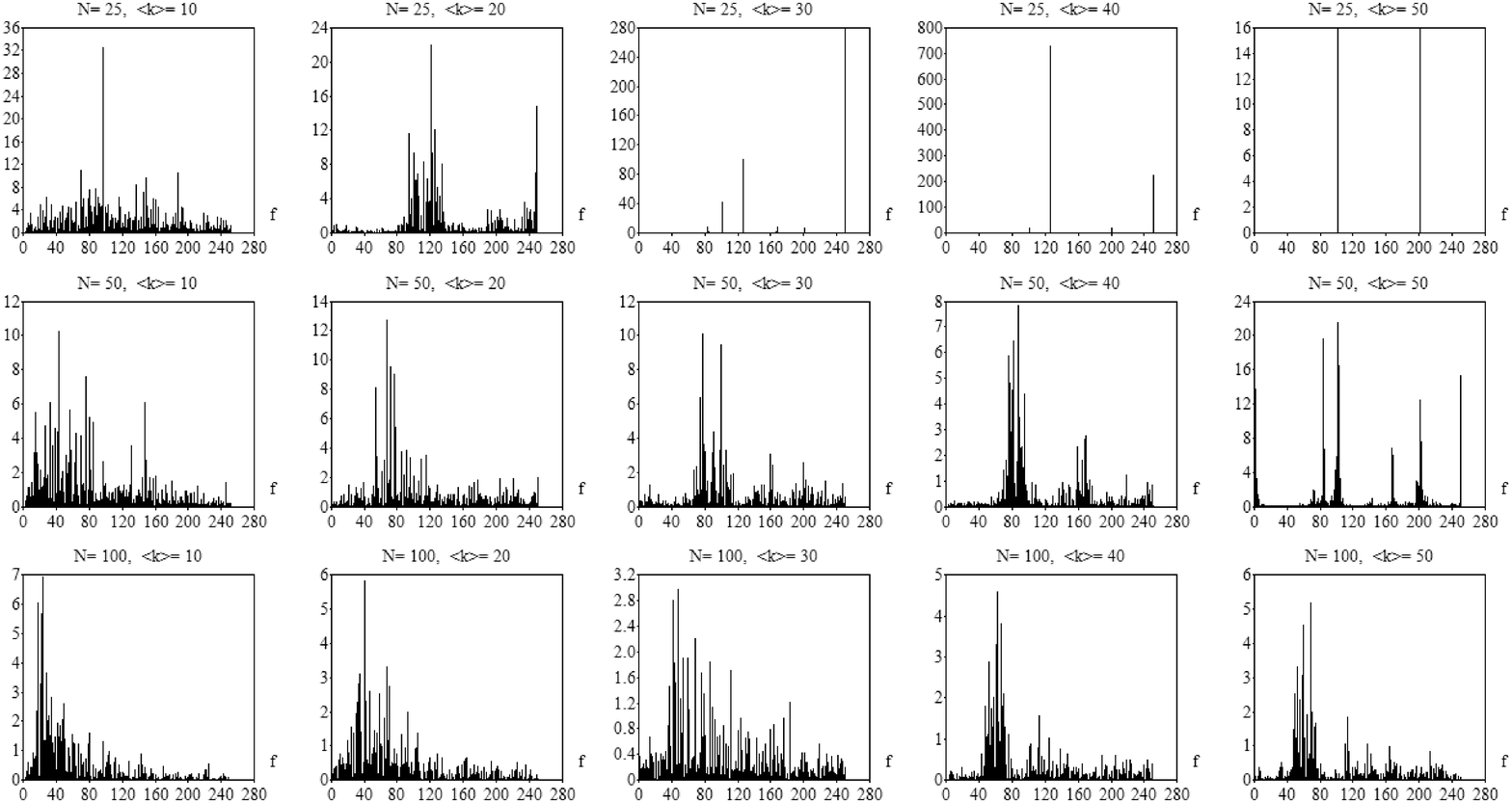} \\ 
  \caption{The
  power spectra obtained for the number of spikes produced by the
  whole original WS networks.  }~\label{fig:spectra_WS} 
 \end{center}
\end{figure*}

Figure~\ref{fig:spectra_WS} depicts the power spectra obtained for
each of the considered WS configurations.  Such spectra, which were
obtained by considering the equilibrium regime of the spikings
(i.e. the time steps from 500 to 1000) are remarkably similar to those
obtained by the ER structures (Fig.~\ref{fig:spectra_ER}, suggesting
some possible universality between different complex networks types.

Regarding the total activation inside the system, the revealed results
(not shown in this work) also included the two regimes as in the ER
case, but with sublinear increase during the transient regime, which
was also duly followed by the dissipative plateau of overall
activation.

Figure~\ref{fig:corrs_ER} shows the scatterplots of the original node
degrees and total number of spikes induced by each individual neuronal
cell.  Though these two features also tended to be organized in terms
of parallel straight groups for small values of average degree, the
correlations are now substantially less definite.

\begin{figure*}[htb]
  \vspace{0.3cm} 
  \begin{center}
  \includegraphics[width=0.9\linewidth]{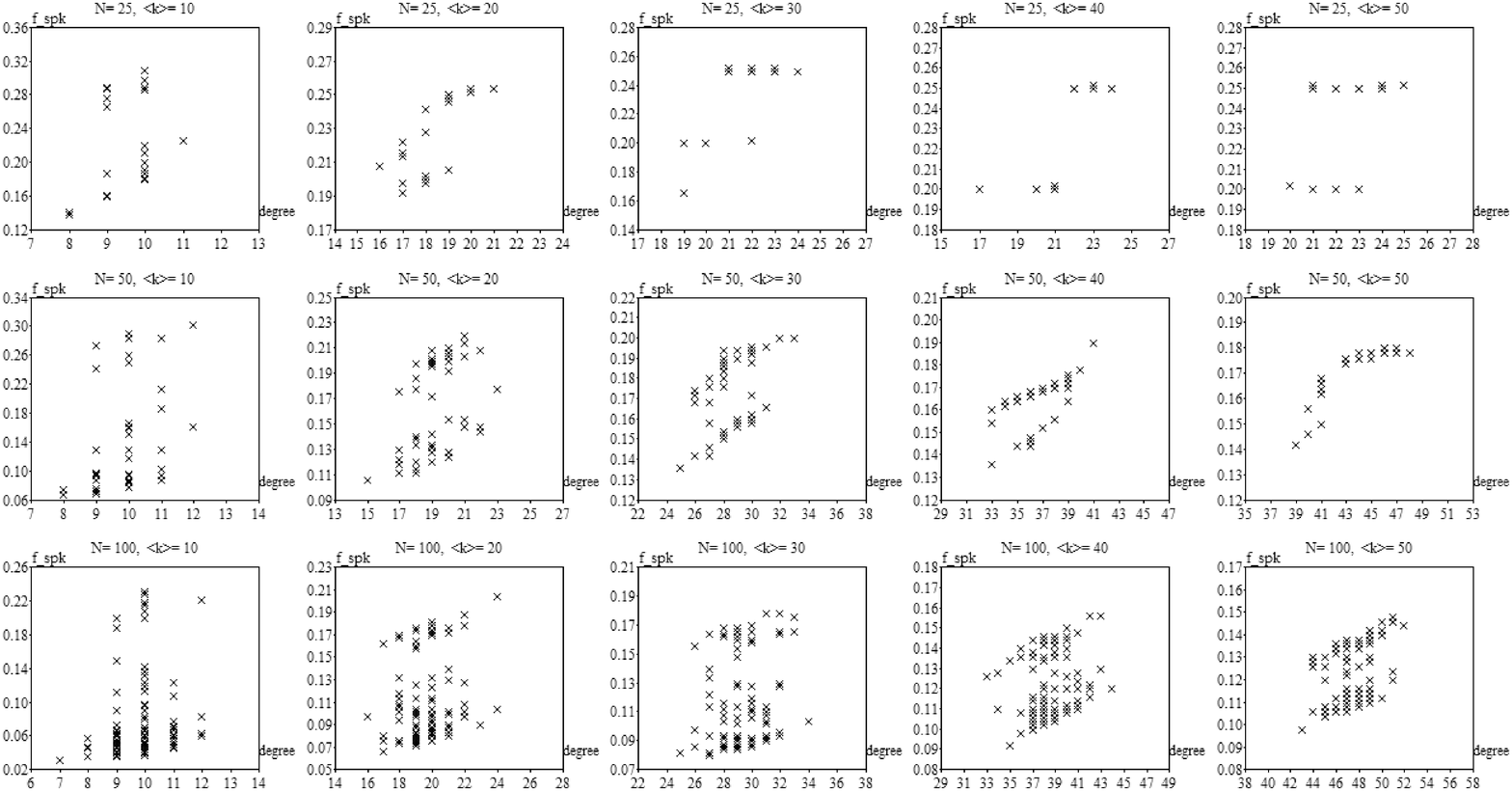} \\  
  \caption{The relationships between the total number of spikes per
             time step
             generated by each neuron and the respective degrees,
             all the considered WS configurations.  The number of
             spikes were counted during 1000 time steps.
  }~\label{fig:corrs_WS} 
  \end{center}
\end{figure*}

\section{The Equivalent Model for Non-Uniform Degrees} 

\begin{figure*}[htb]
  \vspace{0.3cm} 
  \begin{center}
  \includegraphics[width=0.6\linewidth]{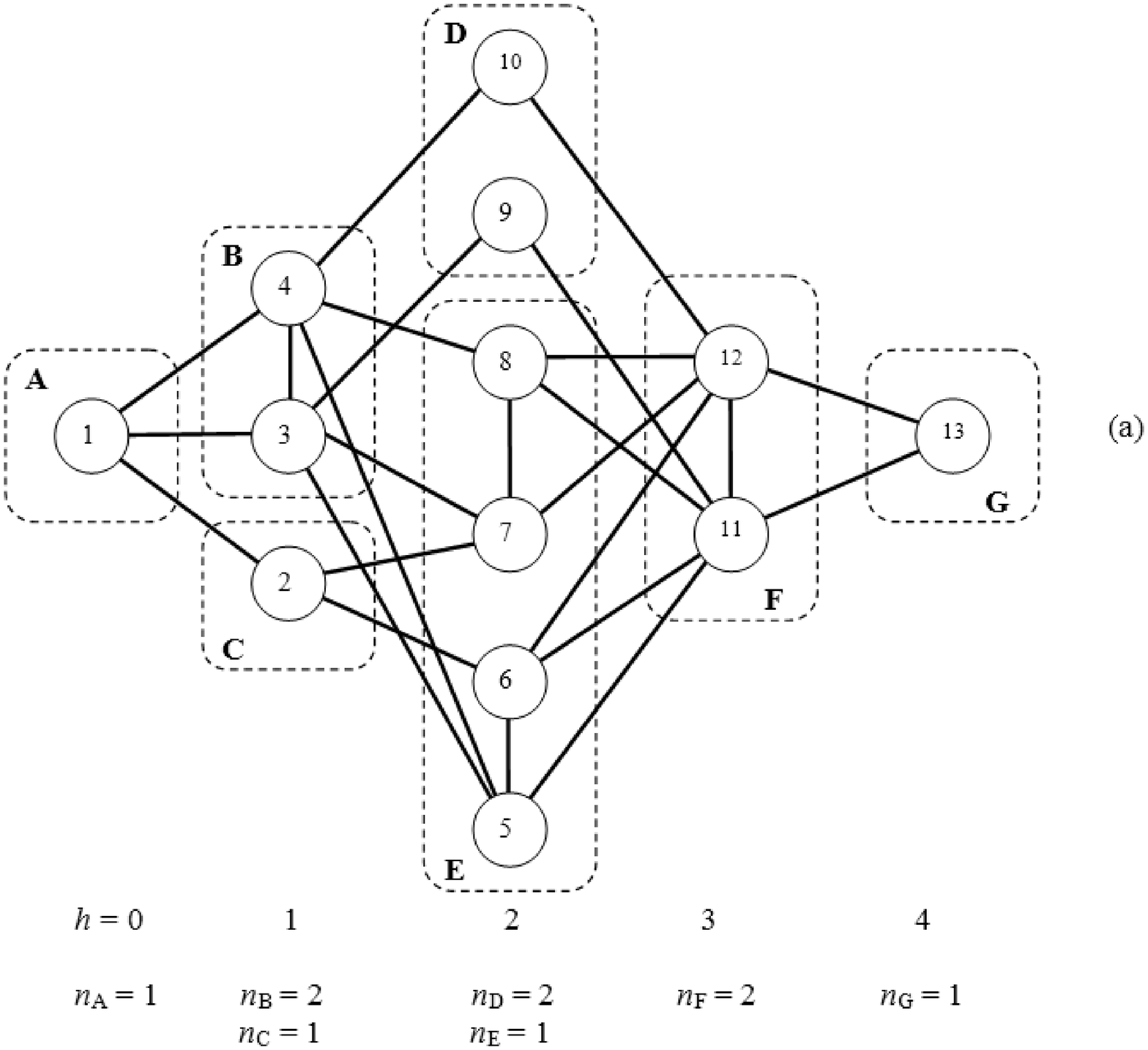} \\  
  \includegraphics[width=0.6\linewidth]{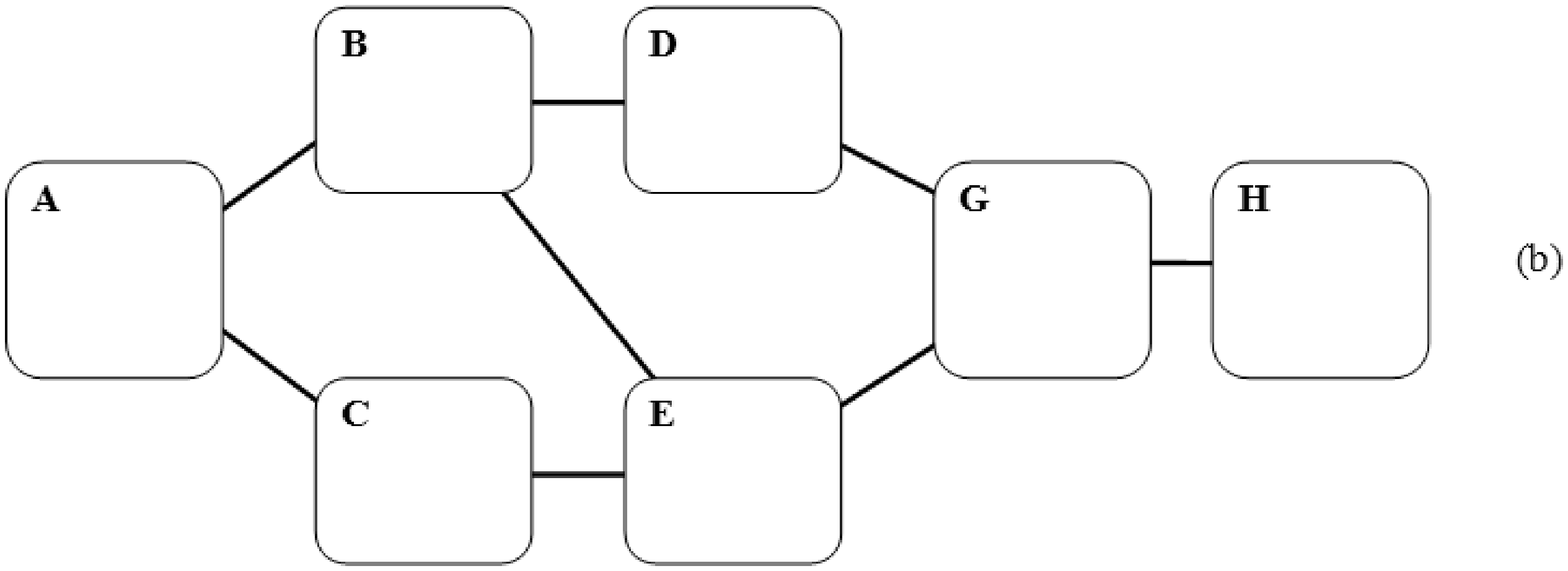} \\  
  \caption{The hierarchical organization of a simple network
              (a) with respect to node 1.  A total of 5 concentric
              levels have been obtained.   The dotted boxed identify the
              nodes with identical degrees at each concentric level,
              which are subsumed by the respective equivalent nodes,
              yielding the equivalent network shown in (b).    
  }~\label{fig:new_model} 
  \end{center}
\end{figure*}

Figure~\ref{fig:new_model} illustrates the procedure which had to be
adopted in order to obtain the a more effective version of the
equivalent model of complex networks, by taking into account
non-uniformities of degrees amongst the original nodes.  First, the
hierarchical organization of the original network with reference to a
given node $i$ is obtained (Fig~\ref{fig:new_model}).  In this
specific case we have 5 concentric levels ($h = 0, 1, \ldots, H=4$),
each containing $1, 3, 6, 2$ and 1 nodes (i.e. $n_0(i)=1$; $n_1(3)=1$;
$n_2(i)=6$; $n_3(i)=2$; $n_4(i)=1$).  The nodes within each concentric
level which have identical degrees are then subsumed into equivalent
nodes.  Thus, nodes 3 and 4 at level $h=1$ --- all with degree equal
to 5 --- become associated to the equivalent node {\bf B}.  Nodes 5, 6
and 7, which have degree 4, are subsumed by node {\bf F}.  The
topology of the resulting equivalent model is illustrated in
Figure~ref{fig:equiv}.

Now, the weights associated to the edges in the equivalent structure
in Figure~\ref{fig:new_model} are determined, similarly as done
in~\cite{Costa_eqcomm:2008}, by means of the following equation,
applied to each pair of equivalent nodes $v$ and $p$ 

\begin{equation}
  W(v,p) = k(v,p)/d  \nonumber \\ \nonumber
\end{equation}

where $k(v,p)$ is the number of original edges going from the nodes
associated to the equivalent node $p$ to the nodes associated to the
equivalent node $v$, $d = \sum_{g \in \Omega(p)} k(g,p)$ and
$\Omega(p)$ is the set of equivalent nodes that receive a directed
edge from $p$. Obseve that $W(b,a)$ expresses the weight of the
connection from $a$ to $b$.  The weights therefore obtained for the
graph in Figure~\ref{fig:new_model} are given as follows

\begin{eqnarray}
  W = \left[
        \begin{array}{ccccccc}
     0   & 1/5 & 1/3 & 0   & 0   & 0    & 0  \\
     2/3 & 1/5 & 0   & 1/2 & 1/4 & 0    & 0  \\
     1/3 & 0   & 0   & 0   & 1/8 & 0    & 0  \\
     0   & 1/5 & 0   & 0   & 0   & 0    & 0  \\
     0   & 2/5 & 0   & 0   & 1/4 & 4/6  & 0  \\
     0   & 0   & 2/3 & 0   & 3/8 & 1/6  & 1  \\
     0   & 0   & 0   & 1/2 & 0   & 1/6  & 0     
      \end{array}  \right]   \nonumber \\ \nonumber
\end{eqnarray}

For instance, as the two original nodes associated to the equivalent
node {\bf B}, both of which with degree 5, send two edges to the
equivalent node {\bf A}, two edges to {\bf D}, four edges to {\bf E},
while two edges remain inside {\bf B}, we have $W(A,B) = 1/5$; $W(D,B)
=1/5$; $W(E,B) =2/5$; $W(B,B)  = 1/5$.

The thresholds associated to neuron corresponding to each equivalent
node are immediately given as the number of original nodes associated
to that equivalent node.  Therefore, we have that $T(A) = 1$; $T(B) =
2$; $T(C) = 1$; $T(D) = 2$; $T(E) = 1$; $T(F) = 3$; $T(G) = 2$; and
$T(H) = 1$.  The limitations to the activations stored into each
memory are also set as being equal to the respective number of nodes
within each equivalent node, i.e. $T(i) = L(i)$.  Observe that the
so-obtained equivalent network itself correspond to an
integrate-and-fire complex neuronal networks with varying thresholds.

\section{Equivalent Model Predictions}

Having characterized the oscillatory integrate-and-fire dynamics along
the transient and equilibrium regimes in terms of several
measurements, it is time to consider the respective equivalent models.
The importance of such an approach lies in the fact that these
structures involve a fraction of the original number of nodes,
therefore allowing the identification and theoretical modeling of more
definite relationships between structure and dynamics.  We start by
considering the distribution of the equivalent nodes and weights, as
well as by comparing the predicted and real number of spikes and
spectra for ER.  

Tables~\ref{tab:n_ER} and~\ref{tab:n_WS} give several features of the
topological structure of the equivalent models obtained for each of
the ER and WS configurations, respectively.  These features include
the total number of original nodes in each concentric level (upper
line), the hierarchical degrees (second line), the total number of
equivalent nodes (in bold, third line), and the number of equivalent
nodes per concentric level (within brackets, third line).  It is clear
from such measurements that the WS structure implied a larger number
of concentric levels in all cases.  On the other hand, most of the ER
structures implied two or three levels (the first of each level
corresponding to the source node), except for the network with $N=100$
and $\left< k \right> = 10$, which implied 4 concentric levels.  Also,
because of its greater degree uniformity, the WS model yielded
substantially fewer equivalent nodes than the ER structures.  The
relatively large number of equivalent nodes obtained for each
concentric level confirms the fact that these two types of supposedly
regular networks do actually a wide dispersion of node degrees,
especially in the case of the ER configurations.  That is the reason
why equivalent models which do not take into account such a
heterogeneity cannot yield accurate predictions.

\begin{table*}
\centering
\begin{tabular}{|c||c|c|c|c|c|}  \hline  
     &  $\left< k \right>=10$ &  $\left< k \right>=20$ &  $\left< k \right>=30$ &  $\left< k \right>=40$ &  $\left< k \right>=50$  \\ \hline
      $N=25$   & 1, 17, 7
               & 1, 23, 1
               & 1, 24 
               & 1, 24
               & 1, 24  \\  
               & 17, 72, 0
               & 23, 23, 0  
               & 24, 0   
               & 24, 0   
               & 24, 0  \\ 
               & {\bf 16:} (1) (10) (5)
               & {\bf 5:} (1) (3) (1)
               & {\bf 2:} (1) (1)
               & {\bf 2:} (1) (1)
               & {\bf 2:} (1) (1)    \\ \hline 
      $N=50$   & 1, 14, 35
               & 1, 35, 14  
               & 1, 42, 7   
               & 1, 45, 4   
               & 1, 49  \\ 
               & 14, 168, 0
               & 35, 310, 0  
               & 42, 246, 0   
               & 45, 176, 0   
               & 49, 0  \\ 
               & {\bf 23:} (1) (10) (12)
               & {\bf 22:} (1) (13) (8)  
               & {\bf 17:} (1) (12) (4)   
               & {\bf 9:} (1) (6) (2)   
               & {\bf 2:} (1) (1)    \\ \hline 
      $N=100$  & 1, 12, 83, 4
               & 1, 36, 63  
               & 1, 67, 32   
               & 1, 68, 31   
               & 1, 75, 24  \\ 
               & 12, 189, 80, 0
               & 36, 859, 0  
               & 67, 1117, 0   
               & 68, 1333, 0   
               & 75, 1358, 0  \\ 
               & {\bf 31:} (1) (9) (17) (4)
               & {\bf 34:} (1) (15) (18)  
               & {\bf 33:} (1) (19) (13)   
               & {\bf 36:} (1) (20) (15)   
               & {\bf 34:} (1) (20) (13)  \\ \hline
\end{tabular}
\caption{Features of the equivalent models obtained for the ER configurations:
                   number of nodes per level (first line); hierarchical
                   degrees of each level (second line); and total number of
                   equivalent nodes (bold, third line) and the number of 
                   equivalent per concentric level (within brackets).
        }\label{tab:n_ER}
\end{table*}

\begin{table*}
\centering
\begin{tabular}{|c||c|c|c|c|c|}  \hline  
     &  $\left< k \right>=10$ &  $\left< k \right>=20$ &  $\left< k \right>=30$ &  $\left< k \right>=40$ &  $\left< k \right>=50$  \\ \hline
      $N=25$   & 1, 10, 12, 2
               & 1, 18, 6  
               & 1, 21, 3   
               & 1, 23, 1   
               & 1, 21, 3  \\ 
               & 10, 32, 15, 0
               & 18, 84, 0  
               & 21, 54, 0   
               & 23, 21, 0   
               & 21, 57, 0  \\ 
               & {\bf 9:} (1) (2) (4) (2)
               & {\bf 10:} (1) (6) (3)   
               & {\bf 9:} (1) (5) (3)   
               & {\bf 8:} (1) (6) (1)   
               & {\bf 10:} (1) (6) (3)  \\ \hline
      $N=50$   & 1, 9, 14, 24, 2
               & 1, 19, 26, 4  
               & 1, 28, 21   
               & 1, 38, 11   
               & 1, 46, 3  \\ 
               & 9, 36, 54, 17, 0
               & 19, 131, 62, 0  
               & 28, 271, 0   
               & 38, 309, 0   
               & 46, 114, 0  \\ 
               & {\bf 15:} (1) (4) (4) (4) (2)
               & {\bf 17:} (1) (6) (8) (2)  
               & {\bf 16:} (1) (8) (7)   
               & {\bf 16:} (1) (9) (6)   
               & {\bf 12:} (1) (8) (3)  \\ \hline  
      $N=100$  & 1, 11, 34, 38, 16
               & 1, 21, 57, 21  
               & 1, 27, 55, 1   
               & 1, 42, 57   
               & 1, 50, 49  \\ 
               & 11, 57, 118, 91, 0 
               & 21, 179, 260, 0  
               & 27, 312, 338, 0   
               & 42, 577, 0   
               & 50, 843, 0  \\ 
               & {\bf 18:} (1) (3) (4) (6) (4)
               & {\bf 21:} (1) (7) (9) (4  
               & {\bf 23:} (1) (8) (9) (5)   
               & {\bf 22:} (1) (11) (10)   
               & {\bf 19:} (1) (9) (9)  \\ \hline
\end{tabular}
\caption{Features of the equivalent models obtained for the WS configurations:
                   number of nodes per level (first line); hierarchical
                   degrees of each level (second line); and total number of
                   equivalent nodes (bold, third line) and the number of 
                   equivalent per concentric level (within brackets).
        }\label{tab:n_WS}
\end{table*}

Though the partitioning of the concentric levels in terms of groups of
nodes with identical degrees was critical for proper modeling of the
oscillatory dynamics, it supplied as byproduct a new hierarchical
measurement corresponding to the number of nodes with specific degrees
at each concentric level.  Because such features have been presently
found to be essential for the integrate-and-fire dynamics, it is
likely that they will also provide valuable resources for the
characterization of the topology of the networks.

\begin{figure*}[htb]
  \vspace{0.3cm} 
  \begin{center}
  \includegraphics[width=0.9\linewidth]{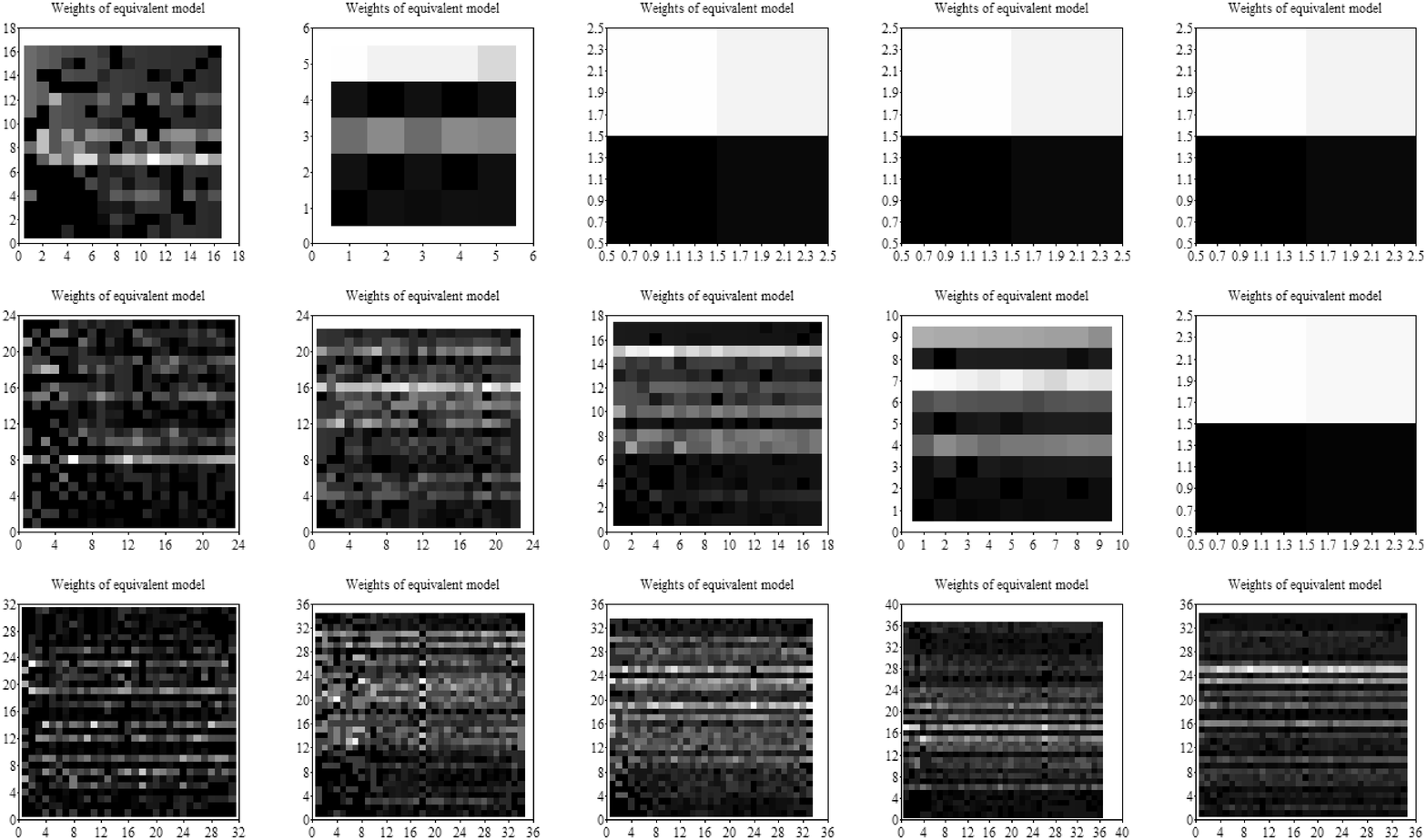} \\  
  \caption{The weights of each of the equivalent models obtained
             for the ER configurations.
  }~\label{fig:weights_ER} 
  \end{center}
\end{figure*}

The weights of the equivalent models obtained for each of the ER
configurations are shown in Figure~\ref{fig:weights_ER}.  The
equivalent nodes are organized bottom-upwards along the columns and
rows according to increasing degrees of the respectively subsumed
original nodes.  It is clear from these matrices that there are some
equivalent nodes which receive converging intense weights (the clearer
rows in the matrices --- recall that we adopt the convention that the
edges extend from the columns towards the rows).  Such nodes are
particularly important for the non-linear dynamics because of their
greater tendency to spike, defining the main avalanches.

\begin{figure*}[htb]
  \vspace{0.3cm} 
  \begin{center}
  \includegraphics[width=0.9\linewidth]{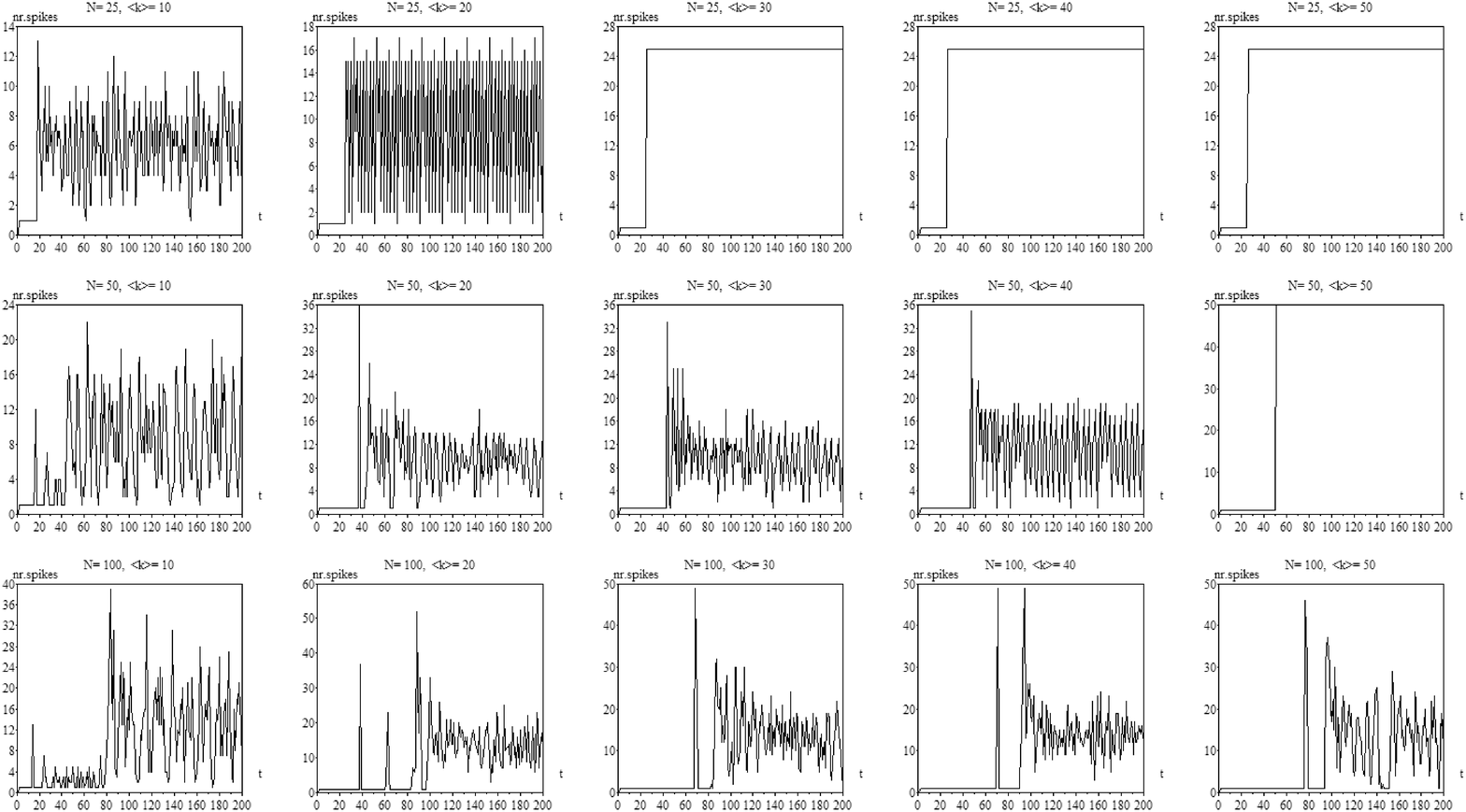} \\  
  \caption{The total number of spikes along time $t$ for the
              ER configurations as predicted
              by the respective equivalent models.
  }~\label{fig:nr_spikes_pred_ER} 
  \end{center}
\end{figure*}

We now present some predictions of the dynamical features as obtained
by using the respective equivalent models.
Figure~\ref{fig:nr_spikes_pred_ER} shows the number of spikes
predicted by the equivalent model for all the considered ER
configurations.  By comparing with the respective real number of
spikes shown in Figure~\ref{fig:nr_spikes_ER}, it becomes clear that
the equivalent model allowed an impressive estimation of both the
transient and steady-state dynamics in every case.

Similarly impressive predictions of several aspects of the dynamics,
including spectra, were also obtained for the WS configurations but
are not shown here.

\section{Equilibrium Configurations: Overall Discussion} \label{sec:discuss}

At this point, after having characterized several aspects of the
integrate-and-fire dynamics obtained for the original networks, as
well as the features of the respective equivalent models and
predictions, it is time to integrate all such information in order to
obtain a more comprehensive explanation of the origin and properties
of the oscillations observed for the integrate-and-fire complex
neuronal networks.

The two clusters of correlations observed for the ER model
(Fig.~\ref{fig:corrs_ER}) are particularly relevant, suggesting that
the spiking frequency tends to increase linearly with the indegree of
the neurons.  Indeed, the higher the indegree, the more activation a
neuron will receive along time, enhancing its chance of firing.
Interestingly, the nodes belonging to each of the two groups straight
correlations in (Fig.~\ref{fig:corrs_ER}) have been found to belong to
distinct respective hierarchical levels.  As the neurons at different
concentric levels tend to fire with different frequencies, depending
on their respective number of nodes and connections, the two groups of
correlations are obtained, with the intra-group straight dispersions
being accounted by the above observed tendency of the spiking
frequency to increase linearly with the indegrees.

Despite the clear separation of the hierarchical levels provided by
the correlation diagrams in Figure~\ref{fig:corrs_ER}, it is
remarkable that the spiking patterns in Figure~\ref{fig:PCA_ER} failed
completely to clusterize with respect to the hierarchies.  This
interesting fact can be taken as an indication that though the nodes
belonging to different hierarchical levels do present a well-defined
\emph{mean frequency}, they are highly irregular as far as the
interspike times are concerned.  This can indeed be corroborated by
the visual analysis of the spikegrams in Figure~\ref{fig:spikes_ER}
as well as from the rich spectra (in the sense of exhibiting many
frequencies) shown in Figure~\ref{fig:spectra_ER}.  With this respect,
it would be interesting to consider PCA projections taking into
account smoothed versions of the spiking patterns, which would enhance
the correlations between those patterns and perhaps make more evident
the hierarchical clusters.

Because the considered WS structures implied larger number of
hierarchical levels (see Tab.~\ref{tab:n_WS}), with respective neurons
firing at distinct group frequencies, the linear relationship between
the spiking frequency and the indegree became blurred in the
respective scatterplots in Figure~\ref{fig:corrs_WS}.

\section{Concluding Remarks}

\subsection{Transient Synchronization}

The subjects of complex networks, neuronal networks and
synchronization have special importance in the investigation of
complex systems and natural phenomena. The current work has brought
these three issues together with respect to transient non-linear
dynamics unfolding in complex networks with different structures and
conservation of activation.  More specifically, the normalized
instantaneous synchronization (NIS) has been proposed as a measurement
of the instantaneous synchronization of the dynamics among the nodes
in the networks at each specific time.  The activation of the networks
was performed by placing a source of unitary activity at each specific
node, and the respective dynamics observed and characterized in terms
of the maximum NIS values, as well as the respective times when they
occurred.

The obtained results indicate that the normalized instantaneous
synchronization tends to increase along the initial steps and then
collapse.  Also, the intrinsic topological organization of each of the
considered types of networks was verified to imply markedly diverse
patterns of activation and maximum NIS.  While relatively uniform
patterns of activation spreading were observed for the ER model,
groups of hubs tended to concentrate the activity in BA networks.
Also, the onset of activation of nodes was verified to be more uniform
for the ER case than for BA.  The WS structure yielded a pattern of
activation which tended to spread gradually amongst the neighbors of
the source.  Similar activation and synchronization were observed for
the two knitted networks, namely PN and PA.  This is particularly
surprising because, though both these networks are defined in terms of
paths, they have completely different degrees of
regularity~\cite{Costa_comp:2007,Costa_longest:2007}.  The different
networks also implied distinct maximum NISs, with the BA resulting
more synchronized along the transient dynamics of activation.  Quite
diverse times of maximum activation were observed for the geographical
network.  Regarding the \emph{C. elegans} network, it was found to
exhibit diverse dynamics with respect to the position of the source.
In addition, most nodes tended to start activity with remarkable
uniformity during the initial 20 time steps.  The overall activation
in this network underwent an abrupt increase after nearly 100 time
steps.  At the longest term, the hubs tended to dominate the
activation dynamics.  It is interesting to observe that the approaches
developed in this work are relevant not only for the synchronization
studies, but also for the characterization of the activation in
non-linear systems underlain by complex connectivity.

The perspectives for future investigations are varied.  Among the
possibilities, it would be interesting to consider other types of
activations, e.g. involving sources at more than one node or periodic
activation instead of the constant values used in this work.  It would
be particularly interesting to study the combined potential of
specific sources for defining diverse dynamical features of the
neuronal activity, especially regarding the facilitation of one source
of activation by other sources.  Another possibility is to consider
the synchronizability of the rates of accesses to a specific node from
activity originating at several nodes~\cite{Costa_sync:2008}.  Such
investigation, which is allowed by the conservation of
activity~\footnote{Interestingly, the activation dynamics in a complex
neuronal network can be thought as involving random walks of moving
agents, so that their movement can be tagged and tracked.}, would
involve the identification of the original source of activations, as
well as its displacement along the networks, during the dissemination
of the activation.  Other interesting questions concern the instant
frequency of spiking along time for each node, as well as the
quantification of correlations and other types of relationships
between the activations.  Because the activation of most of the
considered networks tends to undergo an abrupt dissemination after an
initial transient period, it would be interesting to investigate for
possible critical dynamics (e.g. phase transition).  It would also be
useful to characterize the steady state of activations.  Although the
concepts and methods reported in this article have been considered
from the specific perspective of neuronal networks, they can be
immediately extended to investigations of other situations such as
those involving cortical and biological systems, particularly gene
activation and protein synthesis. 

\subsection{Equilibrium Synchronization}

Though relatively simple, the integrate-and-fire model yields rich
dynamical features, including avalanches, activation confinement
inside communities and oscillations.  Having addressed and explained
the first two phenomena by using equivalent
models~\cite{Costa_eqcomm:2008, Costa_equiv:2008}, it is now
interesting to consider the origins and properties of the oscillations
observed at both transient and equilibrium regimes in
integrate-and-fire complex neuronal networks.  This constituted
precisely the objective of the present work.  Its main contributions
are reviewed and discussed as follows.

{\bf Characterization of Several Aspects of the Oscillations:} The
oscillations in the integrate-and-fire complex neuronal networks were
characterized in terms of several measurements and approaches,
including the visualization of the spikegrams, the total number of
spikes along time and respective power spectra, total activation along
time, PCA decorrelation of the spiking patterns in order to seek for
clusterized dynamics features, as well as correlations between the
total number of spikes and degrees of each neuron.  Each of these
approaches allowed interesting complementary insights about the
oscillatory behavior in uniformly-random (ER) and small-world (WS)
configurations.  Particularly revealing were the twin correlations
observed for both models, which were found to correspond to different
frequencies of oscillations taking place at different concentric
levels of the networks.  Such correlations also indicated that the
mean spiking frequency is linearly related to the indegree of the
respective neurons, which constutes a particularly relevant property
of the analyzed structures.  Analogously to the investigation of
linear diffusive dynamics reported in~\cite{Costa_corrs:2007}, this
property links structure and oscillatory dynamics, implying that
topological hubs will also become hubs of activity in the
integrate-and-fire dynamics.  The scatterplots obtained by the PCA
projection of the individual spiking patterns yielded interesting
structures which, however, were not in correspondence with the
twin-correlation partitioniongs.  The spectra, rich in frequencies,
confirmed the complex structure of the spikings.

{\bf Identification of Two Clearly-Defined Regimes:} The consideration
of the total activation along time revealed two distinct and
well-defined regimes: a conservative transient period, followed by the
dissipative steady-state regime.  The critical point separating these
two dynamics was found to correspond to the main avalanches.  The
transient evolution was found to be more gradual for the WS
configurations, which could be indeed expected because of the less
pronounced (if any) avalanches in this type of networks.

{\bf Equivalent Model for Non-Uniform Degrees:} Following the
comprehensive characterization of the oscillatory properties of the
integrate-and-fire dynamics in complex neuronal networks, an enhanced
equivalent model was developed which takes into account the diversity
of degrees within each concentric topological level.  More
specifically, the nodes with identical degrees at each of the levels
are associated to respective equivalent nodes, allowing a more
detailed model.  Such a modification was found to be critical for
paving the way to impressively accurate predictions of both the
transient and steady-state features of the non-linear dynamics.

{\bf Limitation of the Stored Activation:} Differently from the
previous approaches reported in~\cite{Costa_nrn:2008, Costa_begin:2008,
Costa_activ:2008, Costa_equiv:2008, Costa_eqcomm:2008}, the activation
stored inside the state (memory) of each neuron was limited to be at
most equal to the respective threshold.  In addition to being more
biologically-realistic in which concerns neuronal networks, this
choice also allowed more interesting oscillatory dynamics.

{\bf Additional Hierarchical Measurements of Network Topology:} In
addition to paving the way for precise estimations of the dynamics,
the enhanced equivalent model also yielded as a byproduct new
hierarchical measurements corresponding to the number of nodes with
specific degrees found at each of the respective concentric levels.
Because such topological features proved to be fundamental for
modeling the non-linear integrate-and-fire dynamics, it is expected
that they can also provide valuable features for the characterization
and classification of complex networks~\cite{Costa_surv:2007}.

{\bf Identification of the Origin and Properties of Oscillations:} As
a consequence of the comprehensive characterization and modeling of
the integrate-and-fire dynamics, a better understanding of the origins
and properties of the spiking oscillations have been achieved.  Of
special importance are the positive correlations between the mean
spiking frequency and the neuron indegrees and the fact that nodes at
distinct concentric levels tend to fire with different ensemble
frequencies.  The combination of these two effects yielded the
remarkable twin-correlation diagrams.  In addition, it has been found
that the oscillations tend to unfold after the main avalanches,
i.e. along the steady-state of the dynamics, as revealed by the total
activation in terms of time.

Several are the possibilities for further related investigations.  In
principle, most of the suggestions for future work identified
in~\cite{Costa_nrn:2008, Costa_begin:2008, Costa_activ:2008,
Costa_equiv:2008, Costa_eqcomm:2008} can be immediately extended to
the present work.  More specific possibilities for futher
investigations are identified and briefly discussed in the following.

{\bf Further Simplifications by Considering Degree Intervals:} It
would be interesting to consider the subsuming of the nodes in each
concentric level not strictly under the condition of identical
degrees, but by having similar degrees (i.e. comprised within specific
intervals).  Such a modification would immediately contribute to a
further reduction of the overall number of equivalent nodes, making
the equivalent models more compact.  It would be interesting to
quantify the effect of such a simplification on the accuracy of the
respective predictions of the dynamical features.

{\bf Modular Equivalent Model with Non-Uniform Degrees:} Because of
the finer level of representation allowed by the currently reported
equivalent model, it would be potentially useful to extend the
previous equivalent models used to explain and predict avalanches and
activation confinement inside communities.

{\bf Oscillation Analysis with Action Potentials:} While the current
work, analogously to the previous works~\cite{Costa_nrn:2008,
Costa_begin:2008, Costa_activ:2008, Costa_equiv:2008,
Costa_eqcomm:2008}, considered the activation of each spike to be
equally distributed amongst the outgoing edges (i.e. axons), it would
be interesting to consider the more biologically-realistic hypothesis
that the spikes have constant amplitude (action potential).
Preliminary investigations have already indicated that avalanches,
activation confinement and oscillations are all present in this type
of non-linear dynamics.  However, it would be interesting to perform
more systematic related investigations.

{\bf Chaos and Chaos Control:} The particularly complex spiking
patterns obtained after the main avalanche suggest that the dynamics
of the considered networks may be chaotic.  It would be particularly
promising to apply methods from dynamics systems, such as delay
diagrams and fractal dimensions, in order to search for chaotic
behavior in the integrate-and-fire complex neuronal networks.  Another
interesting perspective would be to apply concepts from chaos control
in order to interfere with the several remarkable aspects of the
investigated dynamics.

{\bf Distinct Weights:} So far, we have been limited to
integrate-and-fire complex neuronal networks containing identical
weights.  Indeed, the weights have only been used in order to obtain
the respective equivalent models.  In this case, the weights
correspond to the proportions of the activations (edges) which are
sent to each equivalent node.  It would be interesting to consider
complex neuronal networks incorporating weights, in order to
investigate the effects of such distributions on the respective
dynamics.

{\bf Investigation of Oscillations in Other Theoretical Complex
Networks Models:} Because of space restrictions, the oscillatory
behavior of integrate-and-fire complex neuronal networks described in
this work has been restricted to a uniformly-random and a small-world
theoretical models of complex networks.  It would be particularly
promising to extend such an investigation to other important types of
network topologies, especially scale free and geographical.  The
consideration of the highly regular knitted
structures~\cite{Costa_comp:2007} would also be interesting.

{\bf Applications to Real-World Problems:} Though we have so far
considered the oscillations in theoretical models of complex neuronal
networks, the respective concepts and methods are immediately
applicable to real-world networks.  Of particular interest would be to
characterize and model the neuronal network of \emph{C. elegans}, as
well as cortical networks.

\begin{acknowledgments}
Luciano da F. Costa thanks CNPq (308231/03-1) and FAPESP (05/00587-5)
for sponsorship.
\end{acknowledgments}

\bibliography{mxd}

\begin{thebibliography}{54}
\expandafter\ifx\csname natexlab\endcsname\relax\def\natexlab#1{#1}\fi
\expandafter\ifx\csname bibnamefont\endcsname\relax
  \def\bibnamefont#1{#1}\fi
\expandafter\ifx\csname bibfnamefont\endcsname\relax
  \def\bibfnamefont#1{#1}\fi
\expandafter\ifx\csname citenamefont\endcsname\relax
  \def\citenamefont#1{#1}\fi
\expandafter\ifx\csname url\endcsname\relax
  \def\url#1{\texttt{#1}}\fi
\expandafter\ifx\csname urlprefix\endcsname\relax\def\urlprefix{URL }\fi
\providecommand{\bibinfo}[2]{#2}
\providecommand{\eprint}[2][]{\url{#2}}

\bibitem[{\citenamefont{Stauffer et~al.}(2003)\citenamefont{Stauffer, Aharony,
  da~F.~Costa, and Adler}}]{Stauffer_Hopfield}
\bibinfo{author}{\bibfnamefont{D.}~\bibnamefont{Stauffer}},
  \bibinfo{author}{\bibfnamefont{L.}~\bibnamefont{Aharony}},
  \bibinfo{author}{\bibfnamefont{L.}~\bibnamefont{da~F.~Costa}},
  \bibnamefont{and} \bibinfo{author}{\bibfnamefont{J.}~\bibnamefont{Adler}},
  \bibinfo{journal}{Eur. Phys. J. B} \textbf{\bibinfo{volume}{32}},
  \bibinfo{pages}{395} (\bibinfo{year}{2003}).

\bibitem[{\citenamefont{da~F.~Costa and Stauffer}(2003)}]{Stauffer_Costa}
\bibinfo{author}{\bibfnamefont{L.}~\bibnamefont{da~F.~Costa}} \bibnamefont{and}
  \bibinfo{author}{\bibfnamefont{D.}~\bibnamefont{Stauffer}},
  \bibinfo{journal}{Physica A} \textbf{\bibinfo{volume}{330}},
  \bibinfo{pages}{37} (\bibinfo{year}{2003}).

\bibitem[{\citenamefont{da~F.~Costa}(2005)}]{Costa_revneur:2005}
\bibinfo{author}{\bibfnamefont{L.}~\bibnamefont{da~F.~Costa}}
  (\bibinfo{year}{2005}), \bibinfo{note}{arXiv:q-bio/0503041}.

\bibitem[{\citenamefont{Kim}(2004)}]{Kim:2004}
\bibinfo{author}{\bibfnamefont{B.~J.} \bibnamefont{Kim}},
  \bibinfo{journal}{Phys. Rev. E} \textbf{\bibinfo{volume}{69}},
  \bibinfo{pages}{045101} (\bibinfo{year}{2004}).

\bibitem[{\citenamefont{Memmesheimer and Timme}(2006)}]{Timme:2006}
\bibinfo{author}{\bibfnamefont{R.~M.} \bibnamefont{Memmesheimer}}
  \bibnamefont{and} \bibinfo{author}{\bibfnamefont{M.}~\bibnamefont{Timme}},
  \bibinfo{journal}{Physica D} \textbf{\bibinfo{volume}{224}},
  \bibinfo{pages}{182} (\bibinfo{year}{2006}).

\bibitem[{\citenamefont{Watts and Strogatz}(1998)}]{Watts_Strogatz:1998}
\bibinfo{author}{\bibfnamefont{D.~J.} \bibnamefont{Watts}} \bibnamefont{and}
  \bibinfo{author}{\bibfnamefont{S.~H.} \bibnamefont{Strogatz}},
  \bibinfo{journal}{Nature} \textbf{\bibinfo{volume}{393}},
  \bibinfo{pages}{409} (\bibinfo{year}{1998}).

\bibitem[{\citenamefont{Watts}(2003)}]{Watts:2003}
\bibinfo{author}{\bibfnamefont{D.}~\bibnamefont{Watts}},
  \emph{\bibinfo{title}{Small Worlds: The Dynamics of Networks between Order
  and Randomness}} (\bibinfo{publisher}{Princeton University Press},
  \bibinfo{year}{2003}).

\bibitem[{\citenamefont{Watts}(2004)}]{Watts:2004}
\bibinfo{author}{\bibfnamefont{D.}~\bibnamefont{Watts}},
  \emph{\bibinfo{title}{Six Degrees: The Science of a Connected Age}}
  (\bibinfo{publisher}{W. W. Norton and Company}, \bibinfo{year}{2004}).

\bibitem[{\citenamefont{Boccaletti et~al.}(2006)\citenamefont{Boccaletti,
  Latora, Moreno, Chavez, and Hwang}}]{Boccaletti:2006}
\bibinfo{author}{\bibfnamefont{S.}~\bibnamefont{Boccaletti}},
  \bibinfo{author}{\bibfnamefont{V.}~\bibnamefont{Latora}},
  \bibinfo{author}{\bibfnamefont{Y.}~\bibnamefont{Moreno}},
  \bibinfo{author}{\bibfnamefont{M.}~\bibnamefont{Chavez}}, \bibnamefont{and}
  \bibinfo{author}{\bibfnamefont{D.}~\bibnamefont{Hwang}},
  \bibinfo{journal}{Phys. Rep.} \textbf{\bibinfo{volume}{424}},
  \bibinfo{pages}{175} (\bibinfo{year}{2006}).

\bibitem[{\citenamefont{Hong et~al.}(2004)\citenamefont{Hong, Kim, Choi, and
  Park}}]{Hong_etal:2004}
\bibinfo{author}{\bibfnamefont{H.}~\bibnamefont{Hong}},
  \bibinfo{author}{\bibfnamefont{B.~J.} \bibnamefont{Kim}},
  \bibinfo{author}{\bibfnamefont{M.~Y.} \bibnamefont{Choi}}, \bibnamefont{and}
  \bibinfo{author}{\bibfnamefont{H.}~\bibnamefont{Park}},
  \bibinfo{journal}{Phys. Rev. E} \textbf{\bibinfo{volume}{69}},
  \bibinfo{pages}{067105} (\bibinfo{year}{2004}).

\bibitem[{\citenamefont{Lee}(2005)}]{Lee:2005}
\bibinfo{author}{\bibfnamefont{D.~S.} \bibnamefont{Lee}},
  \bibinfo{journal}{Phys. Rev. E} \textbf{\bibinfo{volume}{72}},
  \bibinfo{pages}{026208} (\bibinfo{year}{2005}).

\bibitem[{\citenamefont{Hwang et~al.}(2005)\citenamefont{Hwang, Chavez, Amann,
  and Boccaletti}}]{Boccaletti:2005}
\bibinfo{author}{\bibfnamefont{D.~U.} \bibnamefont{Hwang}},
  \bibinfo{author}{\bibfnamefont{M.}~\bibnamefont{Chavez}},
  \bibinfo{author}{\bibfnamefont{A.}~\bibnamefont{Amann}}, \bibnamefont{and}
  \bibinfo{author}{\bibfnamefont{S.}~\bibnamefont{Boccaletti}},
  \bibinfo{journal}{Phys. Rev. letts.} \textbf{\bibinfo{volume}{94}},
  \bibinfo{pages}{138701} (\bibinfo{year}{2005}).

\bibitem[{\citenamefont{Zhou et~al.}(2006)\citenamefont{Zhou, Motter, and
  Kurths}}]{Zhou:2006}
\bibinfo{author}{\bibfnamefont{C.}~\bibnamefont{Zhou}},
  \bibinfo{author}{\bibfnamefont{A.~E.} \bibnamefont{Motter}},
  \bibnamefont{and} \bibinfo{author}{\bibfnamefont{J.}~\bibnamefont{Kurths}},
  \bibinfo{journal}{Phys. Rev. Letts.} \textbf{\bibinfo{volume}{96}},
  \bibinfo{pages}{034101} (\bibinfo{year}{2006}).

\bibitem[{\citenamefont{Boccaletti et~al.}(2007)\citenamefont{Boccaletti,
  Ivachenko, Latora, Pluchino, and Rapisarda}}]{Boccaletti:2007}
\bibinfo{author}{\bibfnamefont{S.}~\bibnamefont{Boccaletti}},
  \bibinfo{author}{\bibfnamefont{M.}~\bibnamefont{Ivachenko}},
  \bibinfo{author}{\bibfnamefont{V.}~\bibnamefont{Latora}},
  \bibinfo{author}{\bibfnamefont{A.}~\bibnamefont{Pluchino}}, \bibnamefont{and}
  \bibinfo{author}{\bibfnamefont{A.}~\bibnamefont{Rapisarda}},
  \bibinfo{journal}{Phys. Rev. E} \textbf{\bibinfo{volume}{75}},
  \bibinfo{pages}{045102} (\bibinfo{year}{2007}).

\bibitem[{\citenamefont{Lodato et~al.}(2007)\citenamefont{Lodato, Boccaletti,
  and Latora}}]{Lodato:2007}
\bibinfo{author}{\bibfnamefont{I.}~\bibnamefont{Lodato}},
  \bibinfo{author}{\bibfnamefont{S.}~\bibnamefont{Boccaletti}},
  \bibnamefont{and} \bibinfo{author}{\bibfnamefont{V.}~\bibnamefont{Latora}},
  \bibinfo{journal}{Phys. Rev. Letts} \textbf{\bibinfo{volume}{78}},
  \bibinfo{pages}{28001} (\bibinfo{year}{2007}).

\bibitem[{\citenamefont{Nishiwaka and Motter}(2006)}]{Takashi:2006}
\bibinfo{author}{\bibfnamefont{T.}~\bibnamefont{Nishiwaka}} \bibnamefont{and}
  \bibinfo{author}{\bibfnamefont{A.~E.} \bibnamefont{Motter}},
  \bibinfo{journal}{Phys. D} \textbf{\bibinfo{volume}{224}},
  \bibinfo{pages}{77} (\bibinfo{year}{2006}).

\bibitem[{\citenamefont{Sorrentino et~al.}(2007)\citenamefont{Sorrentino,
  di~Bernardo, Garofalo, and Chen}}]{Sorrentino:2007}
\bibinfo{author}{\bibfnamefont{F.}~\bibnamefont{Sorrentino}},
  \bibinfo{author}{\bibfnamefont{M.}~\bibnamefont{di~Bernardo}},
  \bibinfo{author}{\bibfnamefont{F.}~\bibnamefont{Garofalo}}, \bibnamefont{and}
  \bibinfo{author}{\bibfnamefont{G.}~\bibnamefont{Chen}},
  \bibinfo{journal}{Phys. Rev. E} \textbf{\bibinfo{volume}{75}},
  \bibinfo{pages}{046103} (\bibinfo{year}{2007}).

\bibitem[{\citenamefont{Sorrentino and Ott}(2007)}]{Ott:2007}
\bibinfo{author}{\bibfnamefont{F.}~\bibnamefont{Sorrentino}} \bibnamefont{and}
  \bibinfo{author}{\bibfnamefont{E.}~\bibnamefont{Ott}},
  \bibinfo{journal}{Phys. Rev. E} \textbf{\bibinfo{volume}{76}},
  \bibinfo{pages}{056114} (\bibinfo{year}{2007}).

\bibitem[{\citenamefont{Almendral and Guilera}(2007)}]{Almendral:2007}
\bibinfo{author}{\bibfnamefont{J.~A.} \bibnamefont{Almendral}}
  \bibnamefont{and} \bibinfo{author}{\bibfnamefont{A.~D.}
  \bibnamefont{Guilera}} (\bibinfo{year}{2007}),
  \bibinfo{note}{arXiv:0705.3216}.

\bibitem[{\citenamefont{da~F.~Costa}(2008{\natexlab{a}})}]{Costa_sync:2008}
\bibinfo{author}{\bibfnamefont{L.}~\bibnamefont{da~F.~Costa}}
  (\bibinfo{year}{2008}{\natexlab{a}}), \bibinfo{note}{arXiv:0801.2520}.

\bibitem[{\citenamefont{Arenas et~al.}(2008)\citenamefont{Arenas, Fernandez,
  and Gomez}}]{Arenas:2008}
\bibinfo{author}{\bibfnamefont{A.}~\bibnamefont{Arenas}},
  \bibinfo{author}{\bibfnamefont{A.}~\bibnamefont{Fernandez}},
  \bibnamefont{and} \bibinfo{author}{\bibfnamefont{S.}~\bibnamefont{Gomez}}
  (\bibinfo{year}{2008}), \bibinfo{note}{arXiv:physics/0703218}.

\bibitem[{\citenamefont{Borisyuk et~al.}(1998)\citenamefont{Borisyuk, Borisyuk,
  and Kazanovich}}]{Borisyuk:1998}
\bibinfo{author}{\bibfnamefont{R.}~\bibnamefont{Borisyuk}},
  \bibinfo{author}{\bibfnamefont{G.}~\bibnamefont{Borisyuk}}, \bibnamefont{and}
  \bibinfo{author}{\bibfnamefont{Y.}~\bibnamefont{Kazanovich}},
  \bibinfo{journal}{Behav. and Brain Sci.} \textbf{\bibinfo{volume}{21}},
  \bibinfo{pages}{833} (\bibinfo{year}{1998}).

\bibitem[{\citenamefont{Aoki and Aoyagi}(2004)}]{Aoki:2004}
\bibinfo{author}{\bibfnamefont{T.}~\bibnamefont{Aoki}} \bibnamefont{and}
  \bibinfo{author}{\bibfnamefont{T.}~\bibnamefont{Aoyagi}}
  (\bibinfo{year}{2004}), \bibinfo{note}{arXiv:q-bio/0410029}.

\bibitem[{\citenamefont{Percha et~al.}(2005)\citenamefont{Percha, Dzakpasu,
  Zochowski, and Parent}}]{Percha:2005}
\bibinfo{author}{\bibfnamefont{B.}~\bibnamefont{Percha}},
  \bibinfo{author}{\bibfnamefont{R.}~\bibnamefont{Dzakpasu}},
  \bibinfo{author}{\bibfnamefont{M.}~\bibnamefont{Zochowski}},
  \bibnamefont{and} \bibinfo{author}{\bibfnamefont{J.}~\bibnamefont{Parent}},
  \bibinfo{journal}{Phys. Rev. E} \textbf{\bibinfo{volume}{72}},
  \bibinfo{pages}{031909} (\bibinfo{year}{2005}).

\bibitem[{\citenamefont{Pereira et~al.}(2007)\citenamefont{Pereira, Baptista,
  and Kurths}}]{Pereira:2007}
\bibinfo{author}{\bibfnamefont{T.}~\bibnamefont{Pereira}},
  \bibinfo{author}{\bibfnamefont{M.~S.} \bibnamefont{Baptista}},
  \bibnamefont{and} \bibinfo{author}{\bibfnamefont{J.}~\bibnamefont{Kurths}}
  (\bibinfo{year}{2007}), \bibinfo{note}{arXiv:0706.3317}.

\bibitem[{\citenamefont{Osipov et~al.}(2007)\citenamefont{Osipov, Kurths, and
  Zhou}}]{Osipov:2007}
\bibinfo{author}{\bibfnamefont{G.~V.} \bibnamefont{Osipov}},
  \bibinfo{author}{\bibfnamefont{J.}~\bibnamefont{Kurths}}, \bibnamefont{and}
  \bibinfo{author}{\bibfnamefont{C.}~\bibnamefont{Zhou}},
  \emph{\bibinfo{title}{Synchronization in Oscillatory Networks}}
  (\bibinfo{publisher}{Springer}, \bibinfo{year}{2007}).

\bibitem[{\citenamefont{Hasegawa}(2004)}]{Hasegawa:2004}
\bibinfo{author}{\bibfnamefont{H.}~\bibnamefont{Hasegawa}},
  \bibinfo{journal}{Phys. Rev. E} \textbf{\bibinfo{volume}{70}},
  \bibinfo{pages}{066107} (\bibinfo{year}{2004}).

\bibitem[{\citenamefont{Hasegawa}(2005)}]{Hasegawa:2005}
\bibinfo{author}{\bibfnamefont{H.}~\bibnamefont{Hasegawa}},
  \bibinfo{journal}{Phys. Rev. E} \textbf{\bibinfo{volume}{72}},
  \bibinfo{pages}{056139} (\bibinfo{year}{2005}).

\bibitem[{\citenamefont{Park and Kim}(2006)}]{Park:2006}
\bibinfo{author}{\bibfnamefont{S.~M.} \bibnamefont{Park}} \bibnamefont{and}
  \bibinfo{author}{\bibfnamefont{B.~J.} \bibnamefont{Kim}},
  \bibinfo{journal}{Phys. Rev. E} \textbf{\bibinfo{volume}{74}},
  \bibinfo{pages}{026114} (\bibinfo{year}{2006}).

\bibitem[{\citenamefont{Latora and Baranger}(1999)}]{Latora_entropy}
\bibinfo{author}{\bibfnamefont{V.}~\bibnamefont{Latora}} \bibnamefont{and}
  \bibinfo{author}{\bibfnamefont{M.}~\bibnamefont{Baranger}},
  \bibinfo{journal}{Physical Review Letters} \textbf{\bibinfo{volume}{82}},
  \bibinfo{pages}{520} (\bibinfo{year}{1999}).

\bibitem[{\citenamefont{Gardenes and Latora}(2007)}]{Gardenes:2007}
\bibinfo{author}{\bibfnamefont{J.~G.} \bibnamefont{Gardenes}} \bibnamefont{and}
  \bibinfo{author}{\bibfnamefont{V.}~\bibnamefont{Latora}}
  (\bibinfo{year}{2007}), \bibinfo{note}{arXiv:0712.0278}.

\bibitem[{\citenamefont{Albert and Barab\'asi}(2002)}]{Albert_Barab:2002}
\bibinfo{author}{\bibfnamefont{R.}~\bibnamefont{Albert}} \bibnamefont{and}
  \bibinfo{author}{\bibfnamefont{A.~L.} \bibnamefont{Barab\'asi}},
  \bibinfo{journal}{Rev. Mod. Phys.} \textbf{\bibinfo{volume}{74}},
  \bibinfo{pages}{47} (\bibinfo{year}{2002}).

\bibitem[{\citenamefont{Newman}(2003)}]{Newman:2003}
\bibinfo{author}{\bibfnamefont{M.~E.~J.} \bibnamefont{Newman}},
  \bibinfo{journal}{SIAM Rev.} \textbf{\bibinfo{volume}{45}},
  \bibinfo{pages}{167} (\bibinfo{year}{2003}).

\bibitem[{\citenamefont{Dorogovtsev and Mendes}(2002)}]{Dorogov_Mendes:2002}
\bibinfo{author}{\bibfnamefont{S.~N.} \bibnamefont{Dorogovtsev}}
  \bibnamefont{and} \bibinfo{author}{\bibfnamefont{J.~F.~F.}
  \bibnamefont{Mendes}}, \bibinfo{journal}{Advs. in Phys.}
  \textbf{\bibinfo{volume}{51}}, \bibinfo{pages}{1079} (\bibinfo{year}{2002}).

\bibitem[{\citenamefont{da~F.~Costa
  et~al.}(2007{\natexlab{a}})\citenamefont{da~F.~Costa, Rodrigues, Travieso,
  and Boas}}]{Costa_surv:2007}
\bibinfo{author}{\bibfnamefont{L.}~\bibnamefont{da~F.~Costa}},
  \bibinfo{author}{\bibfnamefont{F.~A.} \bibnamefont{Rodrigues}},
  \bibinfo{author}{\bibfnamefont{G.}~\bibnamefont{Travieso}}, \bibnamefont{and}
  \bibinfo{author}{\bibfnamefont{P.~R.~V.} \bibnamefont{Boas}},
  \bibinfo{journal}{Advs. in Phys.} \textbf{\bibinfo{volume}{56}},
  \bibinfo{pages}{167} (\bibinfo{year}{2007}{\natexlab{a}}).

\bibitem[{\citenamefont{da~F.~Costa}(2007{\natexlab{a}})}]{Costa_path:2007}
\bibinfo{author}{\bibfnamefont{L.}~\bibnamefont{da~F.~Costa}}
  (\bibinfo{year}{2007}{\natexlab{a}}), \bibinfo{note}{arXiv:0711.1271}.

\bibitem[{\citenamefont{da~F.~Costa}(2007{\natexlab{b}})}]{Costa_comp:2007}
\bibinfo{author}{\bibfnamefont{L.}~\bibnamefont{da~F.~Costa}}
  (\bibinfo{year}{2007}{\natexlab{b}}), \bibinfo{note}{arXiv:0711.2736}.

\bibitem[{\citenamefont{da~F.~Costa}(2007{\natexlab{c}})}]{Costa_longest:2007}
\bibinfo{author}{\bibfnamefont{L.}~\bibnamefont{da~F.~Costa}}
  (\bibinfo{year}{2007}{\natexlab{c}}), \bibinfo{note}{arXiv:0712.0415}.

\bibitem[{\citenamefont{Flory}(1941)}]{Flory}
\bibinfo{author}{\bibfnamefont{P.~J.} \bibnamefont{Flory}},
  \bibinfo{journal}{Journal of the American Chemical Society}
  \textbf{\bibinfo{volume}{63}}, \bibinfo{pages}{3083} (\bibinfo{year}{1941}).

\bibitem[{\citenamefont{da~F.~Costa}(2004{\natexlab{a}})}]{Costa:2004}
\bibinfo{author}{\bibfnamefont{L.}~\bibnamefont{da~F.~Costa}},
  \bibinfo{journal}{Phys. Rev. Lett.} \textbf{\bibinfo{volume}{93}},
  \bibinfo{pages}{098702} (\bibinfo{year}{2004}{\natexlab{a}}).

\bibitem[{\citenamefont{da~F.~Costa}(2004{\natexlab{b}})}]{Costa_perc:2004}
\bibinfo{author}{\bibfnamefont{L.}~\bibnamefont{da~F.~Costa}},
  \bibinfo{journal}{Phys. Rev. E} \textbf{\bibinfo{volume}{70}},
  \bibinfo{pages}{056106} (\bibinfo{year}{2004}{\natexlab{b}}),
  \bibinfo{note}{cond-mat/0312712}.

\bibitem[{\citenamefont{da~F.~Costa and Andrade}(2007)}]{Costa_NJP:2007}
\bibinfo{author}{\bibfnamefont{L.}~\bibnamefont{da~F.~Costa}} \bibnamefont{and}
  \bibinfo{author}{\bibfnamefont{R.~F.~S.} \bibnamefont{Andrade}},
  \bibinfo{journal}{New J. Phys.} \textbf{\bibinfo{volume}{9}},
  \bibinfo{pages}{311} (\bibinfo{year}{2007}).

\bibitem[{\citenamefont{da~F.~Costa and Silva}(2006)}]{Costa_JSP:2006}
\bibinfo{author}{\bibfnamefont{L.}~\bibnamefont{da~F.~Costa}} \bibnamefont{and}
  \bibinfo{author}{\bibfnamefont{F.~N.} \bibnamefont{Silva}},
  \bibinfo{journal}{Journal of Statistical Physics}
  \textbf{\bibinfo{volume}{125}}, \bibinfo{pages}{845} (\bibinfo{year}{2006}).

\bibitem[{\citenamefont{da~F.~Costa and da~Rocha}(2005)}]{Costa_EPJB:2005}
\bibinfo{author}{\bibfnamefont{L.}~\bibnamefont{da~F.~Costa}} \bibnamefont{and}
  \bibinfo{author}{\bibfnamefont{L.~E.~C.} \bibnamefont{da~Rocha}},
  \bibinfo{journal}{The Eur. Phys. J. B} \textbf{\bibinfo{volume}{50}},
  \bibinfo{pages}{237} (\bibinfo{year}{2005}).

\bibitem[{\citenamefont{da~F.~Costa}(2008{\natexlab{b}})}]{Costa_eqcomm:2008}
\bibinfo{author}{\bibfnamefont{L.}~\bibnamefont{da~F.~Costa}}
  (\bibinfo{year}{2008}{\natexlab{b}}), \bibinfo{note}{arXiv:0802.1272}.

\bibitem[{\citenamefont{da~F.~Costa and Cesar}(2001)}]{Costa_book:2001}
\bibinfo{author}{\bibfnamefont{L.}~\bibnamefont{da~F.~Costa}} \bibnamefont{and}
  \bibinfo{author}{\bibfnamefont{R.~M.} \bibnamefont{Cesar}},
  \emph{\bibinfo{title}{Shape Analysis and Classification: {T}heory and
  Practice}} (\bibinfo{publisher}{CRC Press}, \bibinfo{year}{2001}).

\bibitem[{\citenamefont{McLachlan}(1998)}]{McLachlan:1998}
\bibinfo{author}{\bibfnamefont{G.~J.} \bibnamefont{McLachlan}},
  \emph{\bibinfo{title}{Discriminant Analysis and Statistical Pattern
  Recognition}} (\bibinfo{publisher}{John Wiley and Sons},
  \bibinfo{year}{1998}).

\bibitem[{\citenamefont{Duda et~al.}(2000)\citenamefont{Duda, Hart, and
  Stork}}]{Duda_Hart:2000}
\bibinfo{author}{\bibfnamefont{R.~O.} \bibnamefont{Duda}},
  \bibinfo{author}{\bibfnamefont{P.~E.} \bibnamefont{Hart}}, \bibnamefont{and}
  \bibinfo{author}{\bibfnamefont{D.~G.} \bibnamefont{Stork}},
  \emph{\bibinfo{title}{Pattern Classification}} (\bibinfo{publisher}{Wiley
  Interscience}, \bibinfo{year}{2000}).

\bibitem[{\citenamefont{Nicolelis}(1998)}]{Nicolelis:1998}
\bibinfo{author}{\bibfnamefont{M.~A.~L.} \bibnamefont{Nicolelis}},
  \emph{\bibinfo{title}{Methods for Neural Ensemble Recordings}}
  (\bibinfo{publisher}{CRC Press}, \bibinfo{year}{1998}).

\bibitem[{\citenamefont{da~F.~Costa}(2008{\natexlab{c}})}]{Costa_equiv:2008}
\bibinfo{author}{\bibfnamefont{L.}~\bibnamefont{da~F.~Costa}}
  (\bibinfo{year}{2008}{\natexlab{c}}), \bibinfo{note}{arXiv:0802.0421}.

\bibitem[{\citenamefont{da~F.~Costa}(2008{\natexlab{d}})}]{Costa_nrn:2008}
\bibinfo{author}{\bibfnamefont{L.}~\bibnamefont{da~F.~Costa}}
  (\bibinfo{year}{2008}{\natexlab{d}}), \bibinfo{note}{arXiv:0801.3056}.

\bibitem[{\citenamefont{da~F.~Costa}(2008{\natexlab{e}})}]{Costa_begin:2008}
\bibinfo{author}{\bibfnamefont{L.}~\bibnamefont{da~F.~Costa}}
  (\bibinfo{year}{2008}{\natexlab{e}}), \bibinfo{note}{arXiv:0801.4269}.

\bibitem[{\citenamefont{da~F.~Costa}(2008{\natexlab{f}})}]{Costa_activ:2008}
\bibinfo{author}{\bibfnamefont{L.}~\bibnamefont{da~F.~Costa}}
  (\bibinfo{year}{2008}{\natexlab{f}}), \bibinfo{note}{arXiv:0801.4684}.

\bibitem[{\citenamefont{da~F.~Costa
  et~al.}(2007{\natexlab{b}})\citenamefont{da~F.~Costa, Sporns, Antiqueira,
  Nunes, and Oliveira}}]{Costa_corrs:2007}
\bibinfo{author}{\bibfnamefont{L.}~\bibnamefont{da~F.~Costa}},
  \bibinfo{author}{\bibfnamefont{O.}~\bibnamefont{Sporns}},
  \bibinfo{author}{\bibfnamefont{L.}~\bibnamefont{Antiqueira}},
  \bibinfo{author}{\bibfnamefont{M.~G.~V.} \bibnamefont{Nunes}},
  \bibnamefont{and} \bibinfo{author}{\bibfnamefont{O.~N.}
  \bibnamefont{Oliveira}}, \bibinfo{journal}{Appl. Phys. Letts.}
  \textbf{\bibinfo{volume}{91}}, \bibinfo{pages}{054107}
  (\bibinfo{year}{2007}{\natexlab{b}}).

\end{thebibliography}
\end{document}